\documentclass[fleqn]{article}
\pdfoutput=1
\usepackage[margin=1.5in]{geometry} %
\usepackage{mathtools}
\usepackage{amssymb}
\usepackage{mathrsfs}
\usepackage{bbm}
\usepackage{cite}%
\usepackage{graphicx}
\usepackage[table]{xcolor}
\usepackage{nicefrac}
\usepackage[small]{caption2}
\usepackage{cancel}

\newcommand{\rep}[1]{\ensuremath{\boldsymbol{#1}}}
\newcommand{\crep}[1]{\ensuremath{\overline{\boldsymbol{#1}}}}

\newcommand{\I}{\mathrm{i}}

\newcommand{\Z}[1]{\ensuremath{\mathbbm{Z}_{#1}}} 

\newcommand{\x}{\ensuremath{\times}}
\newcommand{\Id}{\mathbbm{1}}

\usepackage[pdftex]{hyperref}
\usepackage{cleveref}

\hypersetup{bookmarksnumbered,colorlinks,
    linkcolor={black},
    citecolor={black},
    urlcolor={black}}

\hypersetup{
    pdftitle = {Eclectic Flavor Groups},
    pdfauthor = {Nilles, Ramos-Sanchez, Vaudrevange}
}

\def\mytitle{Eclectic Flavor Groups}
\title{\mytitle}
\begin{document}
\begin{titlepage}
\vspace*{-3.0cm}

\begin{flushright}
TUM--HEP 1247/19
\end{flushright}

\vspace*{1.0cm}

\begin{center}
{\Large\bfseries\mytitle}

\vspace{1cm}

\textbf{%
Hans Peter Nilles$^{a}$, Sa\'ul Ramos--S\'anchez$^{b,c}$, Patrick K.S. Vaudrevange$^{c}$
}
\\[8mm]
\textit{$^a$\small Bethe Center for Theoretical Physics and Physikalisches Institut der Universit\"at Bonn,\\ Nussallee 12, 53115 Bonn, Germany}
\\[2mm]
\textit{$^b$\small Instituto de F\'isica, Universidad Nacional Aut\'onoma de M\'exico,\\ POB 20-364, Cd.Mx. 01000, M\'exico}
\\[2mm]
\textit{$^c$\small Physik Department T75, Technische Universit\"at M\"unchen,\\ James-Franck-Stra\ss e 1, 85748 Garching, Germany}
\end{center}

\vspace*{1cm}

\begin{abstract}
The simultaneous study of top-down and bottom-up approaches to modular flavor symmetry leads 
necessarily to the concept of eclectic flavor groups. These are nontrivial products of modular and 
traditional flavor symmetries that exhibit the phenomenon of local flavor enhancement in moduli 
space. We develop methods to determine the eclectic flavor groups that can be consistently 
associated with a given traditional flavor symmetry. Applying these methods to a large family of 
prominent traditional flavor symmetries, we try to identify potential candidates for realistic 
eclectic flavor groups and show that they are relatively rare. Model building with finite modular 
flavor symmetries thus appears to be much more restrictive than previously thought.
\end{abstract}
\vspace*{1cm}
\end{titlepage}

\section{Introduction}

While the parameters of the flavor sector of the Standard Model of particle physics have 
essentially all been determined experimentally, the origin of ``flavor'' remains a mystery. Models 
with traditional (discrete) flavor symmetries have provided various fits for masses and mixing angles 
of quarks and leptons~\cite{Feruglio:2019ktm}. More recently, it was suggested that (discrete) 
modular symmetries might describe the flavor structure of the lepton sector~\cite{Feruglio:2017spp}. 
This bottom-up (BU) description is based on the consideration of finite modular groups $\Gamma_N$ 
with $N=2,3,4,5$. Typically, in these models, some of the lepton multiplets are described by nontrivial 
singlets or irreducible triplets of $\Gamma_N$. This suggestion has an important impact on the 
field of lepton flavor physics~\cite{
Altarelli:2005yx,            
deAdelhartToorop:2011re,     
Kobayashi:2018vbk,           
Penedo:2018nmg,              
Criado:2018thu,              
Kobayashi:2018scp,           
Novichkov:2018ovf,           
Novichkov:2018nkm,           
deAnda:2018ecu,              
Okada:2018yrn,               
Kobayashi:2018wkl,           
Novichkov:2018yse,           
Ding:2019xna,                
Nomura:2019jxj,              
Novichkov:2019sqv,           
Okada:2019uoy,               
deMedeirosVarzielas:2019cyj, 
Nomura:2019yft,              
Kobayashi:2019rzp,           
Liu:2019khw,                 
Okada:2019xqk,               
Kobayashi:2019mna,           
Ding:2019zxk,                
Okada:2019mjf,               
King:2019vhv,                
Nomura:2019lnr,              
Okada:2019lzv,               
Criado:2019tzk,              
Kobayashi:2019xvz,           
Asaka:2019vev,               
Chen:2019ewa,                
Gui-JunDing:2019wap,         
Zhang:2019ngf,               
Wang:2019ovr,                
Kobayashi:2019uyt,           
Nomura:2019xsb,              
Kobayashi:2019gtp,           
Lu:2019vgm,                  
Wang:2019xbo                 
}.

To understand the origin of flavor and modular symmetries we need to consider additionally a 
top-down (TD) approach, based on ultraviolet complete theories. Recently, such attempts have 
studied modular symmetries in string derived standard-like models based on heterotic 
orbifolds~\cite{Baur:2019kwi,Baur:2019iai} and magnetized D-branes~\cite{Kobayashi:2016ovu,Kobayashi:2018rad,Kobayashi:2018bff,Kariyazono:2019ehj}
in connection with the standard discrete flavor symmetries within this framework~\cite{Kobayashi:2006wq}. 
This leads to a hybrid picture where the traditional flavor group and the finite modular group 
combine (as a nontrivial product\footnote{In the TD approach, the traditional flavor group does 
not commute with the finite modular group.}) to a generalized flavor group which we call ``Eclectic 
Flavor Group''. It contains the traditional flavor group (which acts universally in moduli space) 
as well as the corresponding modular flavor structure $\Gamma_N$. This picture includes the 
mechanism of ``Local Flavor Unification'' of flavor, $\mathcal{CP}$ and modular symmetries with 
enhanced symmetries at certain locations in moduli space~\cite{Baur:2019kwi,Baur:2019iai}, 
see also ref.~\cite{Gui-JunDing:2019wap}. Furthermore, it potentially incorporates a different flavor 
structure for the quark- and lepton-sector of the Standard Model.

In the present paper we want to analyze possible relations between the bottom-up (BU) and the 
top-down (TD) approaches. At first sight, we are confronted with some potential obstructions. First 
of all, it seems to be difficult to find nontrivial singlets and irreducible triplet 
representations of finite modular groups within the TD approach. Such triplets, as used in the 
BU-case, can be identified more easily for traditional flavor groups, as e.g.\ in $\Delta(54)$ 
of ref.~\cite{Kobayashi:2006wq}. In addition, the TD picture does not always lead to $\Gamma_N$ 
itself (as the finite modular group), but to its double cover ($T'$ in case of $\Gamma_3$, see 
refs.~\cite{Baur:2019kwi,Baur:2019iai}). Although we have studied up to now only a limited number 
of TD-models, we can emphasize the following key observation:
\begin{quote}
The full eclectic flavor group is a nontrivial product of the traditional flavor group, a 
corresponding finite modular group and a $\mathcal{CP}$-like transformation. We cannot treat these 
symmetries separately (as mostly done in the BU-approach) and have to be aware of restrictions (for 
superpotential and K\"ahler potential~\cite{Chen:2019ewa}) from all of these components.
\end{quote}

As a step in our search for a connection between BU- and TD-approaches, we shall develop a 
classification method to obtain all allowed eclectic flavor groups. This is the main goal of the 
present paper. We shall show that this combination of traditional flavor group and finite modular 
group cannot be arbitrary but has to satisfy severe consistency conditions. This can be seen 
already from the results of previous work~\cite{Baur:2019kwi,Baur:2019iai}, where it was 
observed that candidate eclectic flavor groups derive from the traditional flavor group and its 
outer automorphisms.\footnote{This is connected to the outer automorphisms of the Narain space 
group~\cite{GrootNibbelink:2017usl} as explained in refs.~\cite{Baur:2019kwi,Baur:2019iai}.} Based 
on this observation, we shall classify the possible eclectic flavor groups in a bottom-up way 
(for a class of prominent traditional flavor groups) and show that there is only a limited 
number of possibilities.

The paper is structured as follows. In section~\ref{sec:combining} we shall discuss the interplay 
of traditional flavor and modular symmetries and derive relevant consistency conditions. 
Section~\ref{sec:LocalFlavorUnification} discusses the question of ``Local Flavor Unification'' 
from the point of view of allowed modular symmetries. In section~\ref{sec:Delta54Example} we give a 
specific example (closely related to the $\mathbbm{T}^2/\Z{3}$ orbifold) with traditional flavor 
group $\Delta(54)$ and its eclectic extension by $T'$ (being the double cover of $\Gamma_3$). 
Other explicit examples are relegated to the appendices, where we also show that there is no 
eclectic extension of $\Delta(54)$ with $\Gamma_2$ (as a result of the consistency conditions 
derived in section~\ref{sec:combining}). Our results are displayed in Table~\ref{tab:Examples}, 
where one can read off the allowed eclectic flavor groups for representative examples of several
traditional flavor groups. Section~\ref{sec:conclusions} gives conclusions and outlook.

\begin{table}[t!]
\center
\begin{tabular}{|c|cl|cl|cl|cl|}
\hline
 level & \multicolumn{4}{c|}{without $\mathcal{CP}$} & \multicolumn{4}{c|}{with $\mathcal{CP}$} \\
 $N$   & $\Gamma_N$ & GAP       & $\Gamma'_N$          & GAP        & $\Gamma^*_N$     & GAP        & $\Gamma'^*_N$                    & GAP \\
\hline
\hline
2      & $S_3$      & $[ 6, 1]$ & $S_3$                & $[  6, 1]$ & $S_3\times\Z{2}$ & $[ 12, 4]$ & $S_3\times\Z{2}$                 & $[ 12,  4]$ \\
3      & $A_4$      & $[12, 3]$ & $T'$                 & $[ 24, 3]$ & $S_4$            & $[ 24,12]$ & $\mathrm{GL}(2,3)$               & $[ 48, 29]$ \\  
4      & $S_4$      & $[24,12]$ & $\mathrm{SL}(2,4)$   & $[ 48,30]$ & $S_4\times\Z{2}$ & $[ 48,48]$ & $\mathrm{GL}(2,\Z{4})$           & $[ 96,195]$ \\  
5      & $A_5$      & $[60, 5]$ & $\mathrm{SL}(2,5)$   & $[120, 5]$ & $A_5\times\Z{2}$ & $[120,35]$ & $\mathrm{SL}(2,5) \rtimes \Z{2}$ & $[240, 93]$ \\  
\hline
\end{tabular}
\caption{Overview of finite modular groups without and with a $\mathcal{CP}$-like transformation 
(see section~\ref{sec:CP} for the details on $\mathcal{CP}$). The column GAP labels the groups 
according to ref.~\cite{GAP4}, where the first number gives the order of the group. We remark 
several group isomorphisms: $\Gamma_2 \cong\Gamma'_2\cong S_3$, $\Gamma^*_2 \cong\Gamma'^*_2 \cong S_3 \times \Z{2} \cong D_{12}$, 
$\mathrm{SL}(2,2) \cong S_3$ and $\mathrm{SL}(2,3) \cong T'$.}
\label{tab:FiniteModularGroups}
\end{table}

\section{Extending flavor symmetries by modular symmetries}
\label{sec:combining}

The modular group $\mathrm{SL}(2,\mathbbm{Z})$ can be defined by the presentation~\cite{Serre1980}
\begin{equation}\label{eq:SL2Z}
\mathrm{SL}(2,\Z{}) ~=~ \left\langle \mathrm{S}, \mathrm{T} ~|~ \mathrm{S}^4 = \Id,\; 
                       (\mathrm S\mathrm T)^3 = \Id,\; \mathrm S^2\mathrm T = \mathrm T\mathrm S^2 \right\rangle\;,
\end{equation}
and a choice of $\mathrm{SL}(2,\mathbbm{Z})$ generators $\mathrm{S}$ and $\mathrm{T}$ is given by
\begin{equation}\label{eq:SandT}
\mathrm{S} ~=~ \left(\begin{array}{cc} 0 & 1\\ -1 & 0\end{array}\right) \qquad\mathrm{and}\qquad 
\mathrm{T} ~=~ \left(\begin{array}{cc} 1 & 1\\ 0 & 1\end{array}\right)\;.
\end{equation}
Under a modular transformation $\gamma \in \mathrm{SL}(2, \Z{})$ both, the modulus $\tau$ and 
matter fields $\psi$, transform in general nontrivially according to
\begin{equation}\label{eq:ModularTrafoOfFields}
\tau ~\stackrel{\gamma}{\longmapsto}~ \frac{a\,\tau + b}{c\,\tau + d}\;, \qquad 
\psi ~\stackrel{\gamma}{\longmapsto}~ (c\,\tau + d)^{-k} \,\rho(\gamma)\,\psi \quad
    \mathrm{for}\quad \gamma ~=~ \left(\begin{array}{cc} a & b\\ c & d\end{array}\right) ~\in~ \mathrm{SL}(2, \mathbbm{Z})\;.
\end{equation}
Here, $k\in\mathbb{Q}$ is the so-called modular weight\footnote{In contrast to modular forms 
$Y(\tau)$, fields can have fractional weights as realized in string 
theory~\cite{Dixon:1989fj,Ibanez:1992hc,Olguin-Trejo:2017zav}.} of $\psi$ and $\rho(\gamma_1\,\gamma_2)=\rho(\gamma_1)\rho(\gamma_2)$
is a representation of the finite modular group $\Gamma_N$ or of its double cover 
$\Gamma'_N$~\cite{Liu:2019khw} for $N \in\{2,3,4,5\}$. These finite groups are defined by the 
presentations
\begin{subequations}
\label{eq:modularPresentations}
\begin{eqnarray}
  \Gamma_N            & = & \left\langle \mathrm{S}, \mathrm{T} ~|~ \mathrm{S}^2 = \Id,\; (\mathrm S\mathrm T)^3 = \Id,\; \mathrm T^N = \Id \right\rangle\,,\label{eq:modularPresentationsGammaN}\\
  \Gamma'_N           & = & \left\langle \mathrm{S}, \mathrm{T} ~|~ \mathrm{S}^4 = \Id,\; (\mathrm S\mathrm T)^3 = \Id,\; \mathrm T^N = \Id,\; \mathrm S^2\mathrm T = \mathrm T\mathrm S^2 \right\rangle\,.\label{eq:modularPresentationsGammaNPrime}
\end{eqnarray}
\end{subequations}
All finite modular groups $\Gamma_N$ and $\Gamma'_N$ with $N=2,3,4,5$ are listed in 
table~\ref{tab:FiniteModularGroups}. 
\enlargethispage{0.7cm}

The symmetry group of a modular invariant theory is $\mathrm{SL}(2, \Z{})$. This symmetry has 
different realizations for the various fields of the theory: the $\tau$ modulus feels only 
$\mathrm{PSL}(2, \Z{})$ since $\pm\gamma\in\mathrm{SL}(2, \Z{})$ yield the same 
transformation~\eqref{eq:ModularTrafoOfFields} for $\tau$. In contrast, a matter field $\psi$ 
transforms twofold: 
\begin{enumerate}
\item [i)] by the automorphy factor $(c\,\tau + d)^{-k}$ that distinguishes between $\gamma$ and 
$-\gamma$ for general modular weight $k$, and 
\item [ii)] by a linear transformation $\rho(\gamma)$ that can distinguish between $\gamma$ and 
$-\gamma$ only in the case $\Gamma'_N$ with $N > 2$, where $\rho(\gamma)\neq\rho(-\gamma)$ if 
$\rho(-\Id)=\rho(\mathrm{S}^2) \neq \Id$ is nontrivial.
\end{enumerate}
Finally, due to the transformation of the $\tau$ modulus, Yukawa couplings $Y(\tau)$ are in general 
modular forms and transform similar to eq.~\eqref{eq:ModularTrafoOfFields} as 
\begin{equation}\label{eq:ModularTrafoOfYukawas}
Y(\tau) ~\stackrel{\gamma}{\longmapsto}~ Y\left(\frac{a\,\tau + b}{c\,\tau + d}\right) ~=~ (c\,\tau + d)^{k_Y} \,\rho_Y(\gamma)\,Y(\tau) \quad\mathrm{for}\quad \gamma ~=~ \left(\begin{array}{cc} a & b\\ c & d\end{array}\right) ~\in~ \mathrm{SL}(2, \mathbbm{Z})\;,
\end{equation}
where $\rho_Y(\gamma)$ is also a representation of the finite modular group. For a given modular 
weight $k_Y$ (with $k_Y \in 2\mathbb{N}$ for $\Gamma_N$ or $k_Y \in\mathbb{N}$ for $\Gamma'_N$), 
the number of independent Yukawa couplings $Y(\tau)$ is finite and their $\tau$-dependence and 
transformations $\rho_Y(\gamma)$ are explicitly known, see e.g.~\cite{Feruglio:2017spp, Kobayashi:2018vbk, Penedo:2018nmg, Novichkov:2018nkm, Ding:2019xna, Liu:2019khw,Lu:2019vgm}. 
Hence, in order to fully specify a modular invariant theory one has to choose the finite modular 
group $\Gamma_N$ or $\Gamma'_N$ that shall host the representation matrices $\rho(\gamma)$ and 
$\rho_Y(\gamma)$.

Now, since the matrices $\rho(\gamma)$ in eq.~\eqref{eq:ModularTrafoOfFields} and $\rho_Y(\gamma)$ 
in eq.~\eqref{eq:ModularTrafoOfYukawas} must build a (reducible or irreducible) representation of 
$\Gamma_N$ or $\Gamma'_N$, they have to satisfy the respective 
presentation~\eqref{eq:modularPresentationsGammaN} or~\eqref{eq:modularPresentationsGammaNPrime}, 
i.e.\
\begin{equation}\label{eq:GammaN}
\left(\rho(\mathrm{S})\right)^{N_\mathrm{S}} ~=~ \Id\;, \quad \left(\rho(\mathrm{T})\right)^N ~=~ \Id\;, \quad \left(\rho(\mathrm{S}\,\mathrm{T})\right)^3 ~=~ \Id\;, \quad\mathrm{and}\quad \rho(\mathrm{S}^2\mathrm{T}) ~=~ \rho(\mathrm{T}\,\mathrm{S}^2)\;,
\end{equation}
with $N \in\{2,3,4,5\}$ and $N_\mathrm{S} = 2$ for $\Gamma_N$ or $N_\mathrm{S} = 4$ for 
$\Gamma'_N$. 

Let us stress that the $\tau$ modulus transforms nontrivially under modular transformations 
eq.~\eqref{eq:ModularTrafoOfFields}. In contrast to the modular group, we define the traditional 
flavor group $\mathcal{G}_\mathrm{fl}$ by those discrete transformations $g \in \mathcal{G}_\mathrm{fl}$ 
that leave $\tau$ invariant at all points in $\tau$ moduli space, i.e.\ for all $\tau$
\begin{equation}
\tau ~\stackrel{g}{\longmapsto}~ \tau\;, \qquad \psi ~\stackrel{g}{\longmapsto}~ \rho(g)\,\psi \;,
\end{equation}
where $\rho(g)$ is a representation of $\mathcal{G}_\mathrm{fl}$. Hence, Yukawa couplings $Y(\tau)$ 
are invariant under transformations from $\mathcal{G}_\mathrm{fl}$ for all $\tau$.

As we show next, traditional flavor groups are naturally connected to finite modular groups. To 
see this, we apply the modular $\mathrm{S}$ transformation eq.~\eqref{eq:ModularTrafoOfFields} 
twice and obtain
\begin{subequations}
\begin{eqnarray}
\tau & \stackrel{\mathrm{S}}{\longmapsto} & -\frac{1}{\tau} ~\stackrel{\mathrm{S}}{\longmapsto}~ \tau\;,\\
\psi & \stackrel{\mathrm{S}}{\longmapsto} & (-\tau)^{-k} \rho(\mathrm{S})\,\psi ~\stackrel{\mathrm{S}}{\longmapsto}~ \left(\frac{1}{\tau}\right)^{-k}\!\!(-\tau)^{-k}\left(\rho(\mathrm{S})\right)^2\,\psi ~=~ (-1)^{-k}\,\left(\rho(\mathrm{S})\right)^2\,\psi\;.
\end{eqnarray}
\end{subequations}
Since the $\tau$ modulus is invariant under $\mathrm{S}^2$, $\mathrm{S}^2$ is by definition part of 
the traditional flavor group. Moreover, the matter fields $\psi$ transform in general 
nontrivially, $(-1)^{-k}(\rho(\mathrm{S}))^2 \neq \Id$, see also appendix~\ref{app:remarks}. Thus, 
finite modular groups naturally yield traditional flavor groups, and one might wonder how one can 
in general combine a traditional flavor group consistently with a finite modular group.

To answer this question, we derive a constraint on the extension of a traditional flavor group by a 
finite modular group, inspired by the discussions in 
refs.~\cite{Feruglio:2012cw,Holthausen:2012dk,Chen:2014tpa,Novichkov:2019sqv}, where symmetries are 
extended by $\mathcal{CP}$. In detail, we start with a given traditional flavor group 
$\mathcal{G}_\mathrm{fl}$ and try to extend this group consistently by two generators 
$\rho(\mathrm{S})$ and $\rho(\mathrm{T})$ of some finite modular group. To do so, let us consider 
two chains of transformations of the form ``modular, flavor, inverse modular'',
\begin{subequations}\label{eq:ChainOfTransformations}
\begin{eqnarray}
\psi&\stackrel{\mathrm{S}}{\longmapsto}&(-\tau)^{-k}        \,\rho(\mathrm{S})\,\psi~\stackrel{g}{\longmapsto}~(-\tau)^{-k}            \,\rho(\mathrm{S})\,\rho(g)\,\psi ~\stackrel{\mathrm{S}^{-1}}{\longmapsto}~ \rho(\mathrm{S})\,\rho(g)\,\rho(\mathrm{S})^{-1}\,\psi\;,\\
\psi&\stackrel{\mathrm{T}}{\longmapsto}&\phantom{(-\tau)^{-k}}\rho(\mathrm{T})\,\psi~\stackrel{g}{\longmapsto}~\phantom{(-\tau)^{-k}}\,\!\rho(\mathrm{T})\,\rho(g)\,\psi ~\stackrel{\mathrm{T}^{-1}}{\longmapsto}~ \rho(\mathrm{T})\,\rho(g)\,\rho(\mathrm{T})^{-1}\,\psi\;,
\end{eqnarray}
\end{subequations}
for $g\in\mathcal{G}_\mathrm{fl}$. The $\tau$ modulus is invariant under both chains of 
transformations. Since we do not want to enhance the traditional flavor group 
$\mathcal{G}_\mathrm{fl}$ to a larger traditional flavor group $\mathcal{G}'_\mathrm{fl}$ by 
including new generators $\rho(\mathrm{S})\rho(g)\rho(\mathrm{S})^{-1}$ and 
$\rho(\mathrm{T})\rho(g)\rho(\mathrm{T})^{-1}$, we see from eq.~\eqref{eq:ChainOfTransformations} 
that
\begin{equation}\label{eq:ModularFlavorConstraint}
\rho(\mathrm{S})\,\rho(g)\,\rho(\mathrm{S})^{-1} ~\in~ \mathcal{G}_\mathrm{fl} \qquad\mathrm{and}\qquad \rho(\mathrm{T})\,\rho(g)\,\rho(\mathrm{T})^{-1} ~\in~ \mathcal{G}_\mathrm{fl}
\end{equation}
must belong to the traditional flavor group $\mathcal{G}_\mathrm{fl}$ for all 
$g \in \mathcal{G}_\mathrm{fl}$. In other words, due to eq.~\eqref{eq:ModularFlavorConstraint} the 
traditional flavor group $\mathcal{G}_\mathrm{fl}$ must be a normal subgroup of the combined 
group generated by $\rho(\mathrm{S})$, $\rho(\mathrm{T})$ and $\rho(g)$, which we call the eclectic 
flavor group. Moreover, we find 
\begin{equation}\label{eq:FlavorGroupMapped}
\langle \rho(\mathrm{S})\,\rho(g)\,\rho(\mathrm{S})^{-1} ~|~ g ~\in~ \mathcal{G}_\mathrm{fl}\rangle ~\cong~ \langle \rho(\mathrm{T})\,\rho(g)\,\rho(\mathrm{T})^{-1}  ~|~ g ~\in~ \mathcal{G}_\mathrm{fl} \rangle ~\cong~ \mathcal{G}_\mathrm{fl}\;.
\end{equation}
Thus, we can sharpen the constraint~\eqref{eq:ModularFlavorConstraint} as
\begin{equation}\label{eq:ModularFlavorConstraint2}
\rho(\mathrm{S})\,\rho(g)\,\rho(\mathrm{S})^{-1} ~=~ \rho(u_\mathrm{S}(g)) \qquad\mathrm{and}\qquad \rho(\mathrm{T})\,\rho(g)\,\rho(\mathrm{T})^{-1} ~=~ \rho(u_\mathrm{T}(g))\;,
\end{equation}
where due to eq.~\eqref{eq:FlavorGroupMapped} the maps $u_\mathrm{S}$ and $u_\mathrm{T}$ are 
automorphisms of the traditional flavor group $\mathcal{G}_\mathrm{fl}$. Since $\rho(\mathrm{S})$ 
and $\rho(\mathrm{T})$ are assumed to generate a finite modular group, it follows from 
eq.~\eqref{eq:ModularFlavorConstraint2} that the automorphisms $u_\mathrm{S}$ and $u_\mathrm{T}$ 
have to satisfy the defining relations~\eqref{eq:GammaN}, i.e.
\begin{equation}\label{eq:ModularFromAutomorphisms}
\left(u_\mathrm{S}\right)^{N_\mathrm{S}} ~=~ \Id\;, \quad \left(u_\mathrm{T}\right)^N ~=~ \Id\;, \quad \left(u_\mathrm{S}\right)^2  \circ u_\mathrm{T} ~=~ u_\mathrm{T} \circ \left(u_\mathrm{S}\right)^2 \quad\mathrm{and}\quad \left(u_\mathrm{S} \circ u_\mathrm{T}\right)^3 ~=~ \Id\;,
\end{equation}
with $N_\mathrm{S} = 2$ for $\Gamma_N$ and $N_\mathrm{S} = 4$ for $\Gamma'_N$. Note that the 
identity $\Id$ in eq.~\eqref{eq:ModularFromAutomorphisms} has to be understood as the trivial 
automorphism, $\Id\!\!:g \mapsto g$ for all $g\in\mathcal{G}_\mathrm{fl}$, and not as an inner 
automorphism. As a consequence, the finite modular group defined by eq.~\eqref{eq:ModularFromAutomorphisms} 
must be a subgroup of the full automorphism group of the traditional flavor group $\mathcal{G}_\mathrm{fl}$.

Now, we can consider two cases. First, if the traditional flavor group commutes with the finite 
modular group, eq.~\eqref{eq:ModularFlavorConstraint2} yields 
\begin{equation}\label{eq:ModularFlavorConstraint3}
\rho(g) ~=~ \rho(u_\mathrm{S}(g)) \qquad\mathrm{and}\qquad \rho(g) ~=~ \rho(u_\mathrm{T}(g))\;,
\end{equation}
for all $g \in \mathcal{G}_\mathrm{fl}$. Thus, both modular transformations $\mathrm{S}$ and 
$\mathrm{T}$ in eq.~\eqref{eq:ModularFlavorConstraint2} correspond to the trivial automorphism of 
$\mathcal{G}_\mathrm{fl}$, $u_\mathrm{S} = u_\mathrm{T} = \Id$, and the eclectic flavor group is
just given by the direct product extension $\mathcal{G}_\mathrm{fl} \times \Gamma_N$ or 
$\mathcal{G}_\mathrm{fl} \times \Gamma'_N$. In this case, the finite modular group can be chosen 
freely.

Motivated by the TD approach however, we are interested only in the case where the traditional 
flavor group does not commute with the finite modular group. In order to satisfy 
condition~\eqref{eq:ModularFlavorConstraint2} in this case, the traditional flavor group must have 
nontrivial automorphisms as candidates for $u_\mathrm{S}$ and $u_\mathrm{T}$. In general, an 
automorphism can be inner or outer, where for an outer automorphism $u$ there exists no 
$h_u \in \mathcal{G}_\mathrm{fl}$ such that
\begin{equation}
u(g) ~=~ h_u\,g\,h_u^{-1} \quad\mathrm{for\ all}\ g \in \mathcal{G}_\mathrm{fl}\;.
\end{equation}
Thus, we have to decide whether $u_\mathrm{S}$ and $u_\mathrm{T}$ are inner or outer automorphisms 
of the traditional flavor group $\mathcal{G}_\mathrm{fl}$.

Let us first assume that $u_\mathrm{S}$ and $u_\mathrm{T}$ are inner automorphisms of the 
traditional flavor group $\mathcal{G}_\mathrm{fl}$. In this case, the automorphisms 
$u_\mathrm{S}$ and $u_\mathrm{T}$ would be defined as
\begin{align}\label{eq:innerST}
u_\mathrm{S}(g) & ~=~ h_\mathrm{S}\,g\,h_\mathrm{S}^{-1}\;, & 
u_\mathrm{T}(g) & ~=~ h_\mathrm{T}\,g\,h_\mathrm{T}^{-1}\;,\phantom{qqqqqqqqqqqqqqqqqqqqqqqqqqqqq}
\end{align}
for some fixed $h_\mathrm{S}, h_\mathrm{T} \in \mathcal{G}_\mathrm{fl}$. Now, assume that 
$u_\mathrm{S}$ and $u_\mathrm{T}$ satisfy eq.~\eqref{eq:ModularFromAutomorphisms} and, hence, 
generate some finite modular group. Then, the action of the modular generators $\mathrm{S}$ and 
$\mathrm{T}$ can always be compensated by an element of the flavor group. To be specific, consider 
the transformations
\begin{equation}\label{eq:trivialextension}
\psi ~\stackrel{\mathrm{S}}{\longmapsto}~ (-\tau)^{-k} \rho(\mathrm{S})\,\psi ~\stackrel{h_\mathrm{S}^{-1}}{\longmapsto}~ (-\tau)^{-k} \,\Id\, \psi\,,
\end{equation}
and similarly for $\mathrm{T}$, where we used $\rho(\mathrm{S})=\rho(h_\mathrm{S})$ that follows 
from eqs.~\eqref{eq:ModularFlavorConstraint2} and~\eqref{eq:innerST}. 
Hence, the generators of the finite modular group can be redefined 
such that the representation of the finite modular group on matter fields $\psi$ is trivial, 
$\rho(\gamma)=\Id$. Since this group would be a trivial extension, we demand in the following 
that $u_\mathrm{S}$ and $u_\mathrm{T}$ are outer automorphisms.

Once this requirement is met, $u_\mathrm{S}$ and $u_\mathrm{T}$ are subject to 
eqs.~\eqref{eq:ModularFlavorConstraint2} and~\eqref{eq:ModularFromAutomorphisms}, which impose 
strong constraints on the possible extensions of the traditional flavor group by a finite modular 
group.

\subsection[The eclectic extension]{\boldmath The eclectic extension \unboldmath}
\label{sec:eclectic}

From our previous discussion, one can classify for a given traditional flavor group\footnote{See 
e.g.~\cite{Ishimori:2010au} for an extensive list of possible traditional flavor groups.}
$\mathcal{G}_\mathrm{fl}$ all nontrivial extensions by finite modular groups as follows: 
\begin{enumerate}
\item[i)] First, one determines the automorphisms of $\mathcal{G}_\mathrm{fl}$ and chooses two 
particular outer automorphisms $u_\mathrm{S}$ and $u_\mathrm{T}$, 
whose specific properties shall be motivated and explained in detail in the next section in the context of $\mathcal{CP}$.
\item[ii)] Then, one checks whether $u_\mathrm{S}$ and $u_\mathrm{T}$ satisfy 
the presentation of a finite modular group as given in eq.~\eqref{eq:ModularFromAutomorphisms}. 
\item[iii)] Finally, for a given representation $\rho(g)$ of the traditional flavor group, one 
constructs $\rho(\mathrm{S})$ and $\rho(\mathrm{T})$ explicitly using 
eq.~\eqref{eq:ModularFlavorConstraint2} such that $\rho(\mathrm{S})$ and $\rho(\mathrm{T})$ satisfy 
the presentation eq.~\eqref{eq:GammaN} of the same finite modular group as $u_\mathrm{S}$ and 
$u_\mathrm{T}$.
\end{enumerate}
The multiplicative closure of the traditional flavor group $\mathcal{G}_\mathrm{fl}$ and its 
compatible finite modular group $\Gamma_N$ (or $\Gamma'_N)$ is called eclectic flavor group, where 
a potential extension by a $\mathcal{CP}$-like transformation will be discussed in the next section. 
Let us stress that the eclectic flavor group is not a direct product of $\mathcal{G}_\mathrm{fl}$ 
and $\Gamma_N$ (or $\Gamma'_N)$ -- in other words, $\mathcal{G}_\mathrm{fl}$ does not commute with 
$\Gamma_N$ (or $\Gamma'_N)$.

\subsection[Combining with CP]{\boldmath Combining with $\mathcal{CP}$ \unboldmath}
\label{sec:CP}

One can combine the modular group $\mathrm{SL}(2, \mathbbm{Z})$ with a $\mathcal{CP}$-like 
transformation by introducing a new generator $\mathrm{K}_*$, which, on the level of the 
$2 \times 2$ matrices given in eq.~\eqref{eq:SandT}, can be realized as 
\begin{equation}
\mathrm{K}_* ~=~ \left(\begin{array}{cc} 1 & 0\\ 0 & -1\end{array}\right)\;,
\end{equation}
such that $\mathrm{SL}(2, \mathbbm{Z})$ is enhanced to $\mathrm{GL}(2, \mathbbm{Z})$~\cite{Novichkov:2019sqv}.
Under $\mathrm K_*$, the $\tau$ modulus and the matter fields $\psi(x)$ transform as 
\begin{equation}
\label{eq:K*trafo}
\tau ~\stackrel{\mathrm{K}_*}{\longmapsto}~ -\overline{\tau}\;, \qquad \psi(x) ~\stackrel{\mathrm{K}_*}{\longmapsto}~ \rho(\mathrm{K}_*)\,\overline{\psi}(x_P)\;,
\end{equation}
see ref.~\cite{Baur:2019kwi} and also~\cite{Dent:2001cc,Novichkov:2019sqv,Baur:2019iai}. Demanding 
that the $\mathcal{CP}$-like transformation be of order 2, i.e.\ $(\mathrm K_*)^2=\Id$, 
implies\footnote{Note that it is in principle possible that $\mathcal{CP}$ is not of order 
2~\cite{Feruglio:2012cw,Holthausen:2012dk,Chen:2014tpa}. However, in this case $\mathrm{K}_*^2$ 
acts as $\tau \mapsto \tau$ and $\psi(x)\mapsto\rho(K_*)\rho(K_*)^*\psi(x)$, which implies that 
$\rho(K_*)\rho(K_*)^*$ is from the traditional flavor group. As we do not want to extend the 
traditional flavor group by further traditional flavor transformations (like $\rho(K_*)\rho(K_*)^*$), 
we focus on the case $(K_*)^2=\Id$.}
\begin{equation}
  \psi(x) ~\stackrel{\mathrm{K}_*}{\longmapsto}~ \rho(\mathrm{K}_*)\,\overline{\psi}(x_P)
          ~\stackrel{\mathrm{K}_*}{\longmapsto}~ \rho(\mathrm{K}_*)\,\rho(\mathrm{K}_*)^*\,\psi(x)
          ~\stackrel{!}{=}~ \psi(x)
\end{equation}
and therefore
\begin{equation}
\label{eq:Ksquare=1}
  \rho(\mathrm{K}_*)\,\rho(\mathrm{K}_*)^* ~=~ \Id \quad\Leftrightarrow\quad \rho(\mathrm{K}_*)^* = \rho(\mathrm{K}_*)^{-1}\,.
\end{equation}

In general, the additional generator $\rho(\mathrm{K}_*)$ of $\mathcal{CP}$ does not commute with 
$\Gamma_N$ (or $\Gamma_N'$). To see this, let us first consider the chain of transformations
\begin{subequations}\label{eq:ChainOfTransformationsK*SK*}
\begin{eqnarray}
\psi(x) & \stackrel{\mathrm{K}_*}{\longmapsto} & \rho(\mathrm{K}_*)\,\overline\psi(x_P)
         ~\stackrel{\mathrm S}{\longmapsto}~     (-\overline\tau)^{-k}\rho(\mathrm{K}_*)\,\rho(\mathrm{S})^*\,\overline\psi(x_P) \\
        & \stackrel{\mathrm{K}_*}{\longmapsto} & (+\tau)^{-k} \rho(\mathrm{K}_*)\,\rho(\mathrm{S})^*\,\rho(\mathrm{K}_*)^{-1}\,\psi(x)\\
        & \stackrel{!}{=}                      & (+\tau)^{-k} \rho(\gamma)\,\psi(x)\;,\label{eq:ChainOfTransformationsK*SK*Final}
\end{eqnarray}
\end{subequations}
for some modular transformation $\gamma \in \mathrm{SL}(2, \mathbbm{Z})$ that we determine next. 
Under the chain of transformations eq.~\eqref{eq:ChainOfTransformationsK*SK*} the $\tau$ modulus 
transforms as $\tau \mapsto -\nicefrac{1}{\tau}$. Thus, $\gamma = \mathrm{S}$ or $\mathrm{S}^{-1}$ 
from eq.~\eqref{eq:SandT}. Eq.~\eqref{eq:ChainOfTransformationsK*SK*Final} implies that the solution 
is $\gamma = \mathrm{S}^{-1}$. In summary, we have found that 
$\mathrm{K}_*\, \mathrm{S}\, \mathrm{K}_* = \mathrm{S}^{-1}$ on the level of 
$\mathrm{GL}(2, \mathbbm{Z})$. Consequently, the finite modular group has to be extended by 
$\rho(\mathrm{K}_*)$ satisfying
\begin{equation}
\rho(\mathrm{K}_*)\,\rho(\mathrm{S})^*\,\rho(\mathrm{K}_*)^{-1} ~=~ \rho(\mathrm{S})^{-1}\;.
\end{equation}
By repeating these steps for $\mathrm{T}$, we find 
$\mathrm{K}_*\, \mathrm{T}\, \mathrm{K}_* = \mathrm{T}^{-1}$ and, as the final result, that 
the conditions~\eqref{eq:GammaN} get extended by
\begin{subequations}
\label{eq:ModularFlavorConstraintWithK*}
\begin{eqnarray}
\rho(\mathrm{K}_*)^* & = & \rho(\mathrm{K}_*)^{-1}\;,\\
\rho(\mathrm{K}_*)\,\rho(\mathrm{S})^*\,\rho(\mathrm{K}_*)^{-1} & = & \rho(\mathrm{S})^{-1}\;,\\
\rho(\mathrm{K}_*)\,\rho(\mathrm{T})^*\,\rho(\mathrm{K}_*)^{-1} & = & \rho(\mathrm{T})^{-1}\;.
\end{eqnarray}
\end{subequations}
This enhances the finite modular group $\Gamma_N$ to $\Gamma^*_N$ (and enhances $\Gamma'_N$ to 
$\Gamma'^*_N$), defined as
\begin{subequations}
\label{eq:modularCPPresentations}
\begin{eqnarray}
\label{eq:modularCPPresentationsGammaN}
  \Gamma^*_N  & = & \left\langle \mathrm{S}, \mathrm{T}, \mathrm{K}_* ~|~ \mathrm{S}^2 = \Id,\; (\mathrm S\mathrm T)^3 = \Id,\; \mathrm T^N = \Id,\right.\nonumber\\\ 
              &   & \;\left.\mathrm{K}_*^2 = \Id,\; \mathrm{K}_*\, \mathrm{S}\, \mathrm{K}_* = \mathrm{S}^{-1},\;\mathrm{K}_*\, \mathrm{T}\, \mathrm{K}_* = \mathrm{T}^{-1} \right\rangle\,,\label{eq:CPmodularPresentationsGammaN}\\
  \Gamma'^*_N & = & \left\langle \mathrm{S}, \mathrm{T}, \mathrm{K}_* ~|~ \mathrm{S}^4 = \Id,\; (\mathrm S\mathrm T)^3 = \Id,\; \mathrm T^N = \Id,\; \mathrm S^2\mathrm T = \mathrm T\mathrm S^2,\;\right.\nonumber\\
\label{eq:modularCPPresentationsGammaN'}
              &   & \;\left.\mathrm{K}_*^2 = \Id,\; \mathrm{K}_*\, \mathrm{S}\, \mathrm{K}_* = \mathrm{S}^{-1},\;\mathrm{K}_*\, \mathrm{T}\, \mathrm{K}_* = \mathrm{T}^{-1} \right\rangle\,.\label{eq:CPmodularPresentationsGammaNPrime}
\end{eqnarray}
\end{subequations}
All finite modular groups with a $\mathcal{CP}$-like extension and $N=2,3,4,5$ are listed in 
table~\ref{tab:FiniteModularGroups}.

Next, we discuss how $\Gamma^*_N$ can be made compatible with the traditional flavor group 
$\mathcal{G}_\mathrm{fl}$, cf.~\cite{Feruglio:2012cw,Holthausen:2012dk,Chen:2014tpa}. With the 
additional element $\mathrm K_*$, our previous discussion proceeds directly. First, 
eq.~\eqref{eq:ChainOfTransformations} includes now also the chain of transformations
\begin{equation}
\label{eq:ChainOfTransformationsK*}
\psi(x) \stackrel{\mathrm{K}_*}{\longmapsto}~\rho(\mathrm{K}_*)\,\overline\psi(x_P)
   ~\stackrel{g}{\longmapsto}~\rho(\mathrm{K}_*)\,\rho(g)^*\,\overline\psi(x_P) 
   ~\stackrel{\mathrm{K}_*^{-1}}{\longmapsto}~ \rho(\mathrm{K}_*)\,\rho(g)^*\,\rho(\mathrm{K}_*)^{-1}\,\psi(x)\;,
\end{equation}
which implies $\rho(\mathrm{K}_*)\,\rho(g)^*\,\rho(\mathrm{K}_*)^{-1}\in\mathcal{G}_\mathrm{fl}$ 
when we prevent $\mathcal{G}_\mathrm{fl}$ from being trivially extended by the elements 
$\rho(\mathrm{K}_*)\,\rho(g)^*\,\rho(\mathrm{K}_*)^{-1}$. It then follows that there exists an 
automorphism $u_{\mathrm{K}_*}$, such that
\begin{equation}
 \label{eq:Definition-uK*}
 \rho(\mathrm{K}_*)\rho(g)^*\rho(\mathrm{K}_*)^{-1} = \rho(u_{\mathrm{K}_*}(g))\;.
\end{equation}
As shown in ref.~\cite{Chen:2014tpa}, eq.~\eqref{eq:Definition-uK*} is satisfied
by a class-inverting outer automorphism of the traditional flavor group $\mathcal{G}_\mathrm{fl}$.
However, one can also have a situation in which not all irreducible representations of $\mathcal{G}_\mathrm{fl}$
appear in the theory and there exists an automorphism $u_{\mathrm{K}_*}$ satisfying
eq.~\eqref{eq:Definition-uK*} only for the representation(s) $\rho$ present in the spectrum. 
Such an automorphism could then be seen as a $\rho$-restricted class-inverting 
automorphism.\footnote{We thank Andreas Trautner for useful comments on the properties of
class-inverting automorphisms.}
Now, since this type of outer automorphisms necessarily doubles the dimensions of (some of) 
the representations and because in the TD approach they give rise to $\mathcal{CP}$-like 
transformations~\cite{Baur:2019kwi,Baur:2019iai}, we reserve these $\rho$-restricted class-inverting 
automorphisms of $\mathcal{G}_\mathrm{fl}$ exclusively for $\mathcal{CP}$.

Consequently, the automorphisms $u_{\mathrm{K}_*}$, $u_\mathrm{S}$ and $u_\mathrm{T}$ of the 
traditional flavor group $\mathcal{G}_\mathrm{fl}$ have to satisfy (in addition to 
eq.~\eqref{eq:ModularFromAutomorphisms}) the conditions
\begin{equation} \label{eq:ModularFromAutomorphismsWithK*}
 (u_{\mathrm{K}_*})^2~=~\Id\,,\quad u_{\mathrm{K}_*}\circ u_\mathrm{S}\circ u_{\mathrm{K}_*}~=~u_{\mathrm{S}}^{-1}\,,\quad
  u_{\mathrm{K}_*}\circ u_\mathrm{T}\circ u_{\mathrm{K}_*}~=~u_{\mathrm{T}}^{-1}\;.
\end{equation}
In more detail, by applying the definitions of the automorphisms $u_\mathrm{S}$ and 
$u_{\mathrm{K}_*}$ in eqs.~\eqref{eq:ModularFlavorConstraint2} and~\eqref{eq:Definition-uK*}, the 
second relation is obtained as follows:
\begin{eqnarray}
 \rho\left(u_{\mathrm{K}_*}\circ u_\mathrm{S}\circ u_{\mathrm{K}_*}(g)\right)  
      &=& \rho(\mathrm{K}_*)\rho\left(u_\mathrm{S}\circ u_{\mathrm{K}_*}(g)\right)^*\rho(\mathrm{K}_*)^{-1} \nonumber\\
      &=& \rho(\mathrm{K}_*)\rho(\mathrm{S})^*\rho\left(u_{\mathrm{K}_*}(g)\right)^*{\rho(\mathrm{S})^*}^{-1}\rho(\mathrm{K}_*)^{-1} \nonumber\\
      &=& \rho(\mathrm{K}_*)\rho(\mathrm{S})^*\rho(\mathrm{K}_*)^{-1}\rho(g)\rho(\mathrm{K}_*){\rho(\mathrm{S})^*}^{-1}\rho(\mathrm{K}_*)^{-1} \nonumber\\  
      &=& \rho(\mathrm{S})^{-1}\rho(g)\rho(\mathrm{S}) ~=~ \rho(u_\mathrm{S}^{-1}(g))\,.
\end{eqnarray}

As before, for each traditional flavor group $\mathcal{G}_\mathrm{fl}$, it is possible to classify 
all finite modular groups $\Gamma_N^*$ (and $\Gamma'^*_N$), endowed with a $\mathcal{CP}$-like 
transformation, that are compatible with $\mathcal{G}_\mathrm{fl}$. One must first create all 
automorphisms of $\mathcal G_\mathrm{fl}$ and choose outer automorphisms 
$u_\mathrm{S}$, $u_\mathrm{T}$ and $u_{\mathrm{K}_*}$ 
satisfying the presentation of an enhanced finite modular group given by 
eqs.~\eqref{eq:ModularFromAutomorphisms} and~\eqref{eq:ModularFromAutomorphismsWithK*}. Finally, 
one must explicitly find the representations $\rho(\mathrm S)$, $\rho(\mathrm T)$ and 
$\rho(\mathrm K_*)$ that fulfill eqs.~\eqref{eq:ModularFlavorConstraint2},~\eqref{eq:ModularFlavorConstraintWithK*} 
and~\eqref{eq:Definition-uK*}.

Let us make a remark on the non-Abelian structure of the $\mathcal{CP}$-like extension of finite 
modular groups. Despite the fact that eq.~\eqref{eq:modularCPPresentationsGammaN} indicates that, 
for $N>2$, the generator $\mathrm K_*$ of $\mathcal{CP}$ does not commute with $\mathrm S$ and 
$\mathrm T$, the $\mathcal{CP}$-like extension of $\Gamma_N$ becomes $\Gamma_N^* = \Gamma_N\x\Z{2}$ 
for several finite modular groups. In detail, the $\Z{2}$ factors in the $\Gamma^*_{N}$ finite 
modular groups $S_3\times\Z{2}$, $S_4\times\Z{2}$ and $A_5\times\Z{2}$ are generated by 
$\mathrm{K}_*$, $(\mathrm{T}^2\mathrm{S})^2\mathrm{K}_*$ and 
$\mathrm{S}(\mathrm{T}^2\mathrm{S}\mathrm{T})^2\mathrm{T}\mathrm{K}_*$, respectively, see 
table~\ref{tab:FiniteModularGroups}.

\section{Local Flavor Unification}
\label{sec:LocalFlavorUnification}

We are considering a setting, where modular and traditional flavor symmetries do not commute. This 
gives rise to the picture of ``Local Flavor Unification''~\cite{Baur:2019kwi,Baur:2019iai}: At 
so-called self-dual points or lines in moduli space $\langle \tau\rangle$ the finite modular symmetry 
is broken spontaneously to those subgroups that leave $\langle \tau\rangle$ invariant. In contrast, 
the traditional flavor symmetry, by definition, leaves the modulus $\tau$ invariant and, hence, remains 
unbroken everywhere in moduli space. As modular and traditional flavor symmetries do not commute, 
the unbroken modular transformations yield nontrivial enhancements of the traditional flavor 
symmetry to the so-called unified flavor symmetries at the self-dual points in moduli space.

Let us begin the discussion with the finite modular group $\Gamma_N$ or $\Gamma'_N$, i.e.\ without 
taking $\mathcal{CP}$-like transformations into account. Then, if the modulus is stabilized at 
$\langle \tau\rangle = \I$, the following modular transformation remains unbroken
\begin{equation}\label{eq:SymEnhancementAtI}
\langle \tau\rangle ~\stackrel{\mathrm{S}}{\longmapsto}~ -\frac{1}{\langle \tau\rangle} ~=~ \langle \tau\rangle \qquad\mathrm{at}\quad \langle \tau\rangle ~=~ \I\;,
\end{equation}
see also the related discussion in ref.~\cite{Gui-JunDing:2019wap}. At this point in moduli space, 
matter fields transform as
\begin{equation}
\psi ~\stackrel{g}{\longmapsto}~ \rho(g)\,\psi \quad\mathrm{for}\quad g~\in~ \mathcal{G}_\mathrm{fl} \qquad\mathrm{and}\qquad \psi ~\stackrel{g}{\longmapsto}~ (-\I)^{-k}\rho(g)\,\psi \quad\mathrm{for}\quad g ~=~ \mathrm{S}\;.
\end{equation}
Thus, at $\langle \tau\rangle = \I$ the traditional flavor symmetry $\mathcal{G}_\mathrm{fl}$ is 
enhanced by the generator $\mathrm{S}$. Since $\rho(\mathrm{S})$ is either of order 
$N_\mathrm{S}=2$ or $N_\mathrm{S}=4$, see eq.~\eqref{eq:GammaN}, the unified flavor 
symmetry at $\langle \tau\rangle = \I$ is a nontrivial extension of $\mathcal{G}_\mathrm{fl}$ by 
$\Z{N_\mathrm{S}}$~\cite{Baur:2019kwi,Baur:2019iai}, or even of higher order depending on the 
modular weight $k\in\mathbb{Q}$.

If the modulus is stabilized at $\langle \tau\rangle = \exp\left(\nicefrac{2\pi\I}{3}\right)$, we 
find the following unbroken modular transformation
\begin{equation}
\langle \tau\rangle ~\stackrel{\mathrm{S}\,\mathrm{T}}{\longmapsto}~ -\frac{1}{\langle \tau\rangle + 1} ~=~ \langle \tau\rangle \qquad\mathrm{at}\quad \langle \tau\rangle ~=~ \exp\left(\nicefrac{2\pi\I}{3}\right) \quad\mathrm{for}\quad \mathrm{S}\,\mathrm{T} ~=~ \left(\begin{array}{cc} 0 & 1\\-1 &-1\end{array}\right)\;.
\end{equation}
Thus, at $\langle \tau\rangle = \exp\left(\nicefrac{2\pi\I}{3}\right)$ the traditional flavor 
symmetry is enhanced by $\mathrm{S}\,\mathrm{T}$ and matter fields transform as 
\begin{equation}
\psi ~\stackrel{g}{\longmapsto}~ \rho(g)\,\psi \quad\mathrm{for}\quad g~\in~ \mathcal{G}_\mathrm{fl} 
\quad\mathrm{and}\quad \psi ~\stackrel{g}{\longmapsto}~ \exp\left(\nicefrac{2\pi\I k}{3}\right)\,\rho(g)\,\psi 
\quad\mathrm{for}\quad g ~=~ \mathrm{S}\,\mathrm{T}
\end{equation}
under the unified flavor symmetry at $\langle \tau\rangle = \exp\left(\nicefrac{2\pi\I}{3}\right)$. 
Note that $\rho(\mathrm{S}\,\mathrm{T}) = \rho(\mathrm{S})\rho(\mathrm{T})$ is of order 3, see 
eq.~\eqref{eq:GammaN}. Thus, the unified flavor symmetry at 
$\langle \tau\rangle = \exp\left(\nicefrac{2\pi\I}{3}\right)$ is a nontrivial product of 
$\mathcal{G}_\mathrm{fl}$ and a $\Z{3}$ generated by $\mathrm{S}\,\mathrm{T}$ (if $k \in\mathbb{Z}$).

If one includes the $\mathcal{CP}$-like transformation, additional modular transformations can 
remain unbroken. For example, at vertical lines in moduli space given by 
$\langle \tau\rangle = \nicefrac{n_{\!B}}{2}+\I\nicefrac{\sqrt{3}r}{2}$ with $n_{\!B} \in \mathbb{Z}$ 
and $r \in \mathbb{R}$ we find that $\mathrm{T}^{n_{\!B}}\,\mathrm{K}_*$ remains 
unbroken~\cite{Baur:2019kwi,Baur:2019iai}, i.e.
\begin{equation}
\langle \tau\rangle ~\stackrel{\mathrm{T}^{n_{\!B}}\,\mathrm{K}_*}{\longmapsto}~ n_{\!B} - \overline{\langle \tau\rangle} ~=~ \langle \tau\rangle 
\quad\mathrm{at}\quad \langle \tau\rangle ~=~ \frac{n_{\!B}}{2}+\I\frac{\sqrt{3}}{2}r
\;\;\;\;\mathrm{for}\;\; \mathrm{T}^{n_{\!B}}\,\mathrm{K}_* ~=~ \left(\begin{array}{cc} 1 & -n_{\!B}\\0 &-1\end{array}\right)\;.
\end{equation}
In particular, for $n_{\!B}=0$, i.e.\ on the vertical line $\langle \tau\rangle = \I\nicefrac{\sqrt{3}r}{2}$ 
we obtain an unbroken modular transformation $\mathrm{K}_*$, while at $\langle \tau\rangle = \I$ 
eq.~\eqref{eq:SymEnhancementAtI} yields an unbroken $\mathrm{S}$ transformation. 
Moreover, if one moves away from $\langle \tau\rangle = \I$ but stays on the circle 
$\langle \tau\rangle = \exp(\I\alpha)$, only the combined transformation $\mathrm{K}_*\mathrm{S}$ 
remains unbroken
\begin{equation}
\langle \tau\rangle ~\stackrel{\mathrm{K}_*\mathrm{S}}{\longmapsto}~ \frac{1}{\overline{\langle \tau\rangle}}  = \langle\tau\rangle 
\quad\mathrm{at}\quad \langle \tau\rangle = \exp(\I\alpha)
\quad\mathrm{for}\quad \mathrm{K}_*\mathrm{S} ~=~ \left(\begin{array}{cc} 0 & 1\\1 & 0\end{array}\right)\;.
\end{equation}
At these lines in moduli space, a $\mathcal{CP}$-like transformation is unbroken. However, it is 
easy to break $\mathcal{CP}$ spontaneously by moving $\langle \tau\rangle$ away from these symmetry 
enhanced lines in moduli space.

\section[Example: Delta(54) and modular symmetries]{\boldmath Example: $\Delta(54)$ and modular symmetries\unboldmath}
\label{sec:Delta54Example}

As an example, let us discuss the possible extensions of the traditional flavor symmetry 
$\Delta(54)$~\cite{Escobar:2008vc,Lam:2008sh,Ishimori:2008uc,Grimus:2008vg}. $\Delta(54)$ is a 
non-Abelian group of order 54 that has three-dimensional representations suitable for the three 
generations of quarks and leptons. In detail, $\Delta(54)$ can be generated by three generators 
$\mathrm{A}$, $\mathrm{B}$, and $\mathrm{C}$ subject to the presentation 
\begin{equation}
\Delta(54) ~=~ \langle \mathrm{A}, \mathrm{B},\mathrm{C} ~|~ \mathrm{A}^3 = \mathrm{B}^3 = \mathrm{C}^2 = (\mathrm{A}\mathrm{C})^2 = (\mathrm{B}\mathrm{C})^2 = (\mathrm{A}\mathrm{B})^3 = (\mathrm{A}\mathrm{B}^2)^3 = \Id \rangle\;.
\end{equation}
A three-dimensional representation of $\Delta(54)$ is given by
\begin{equation}\label{eq:ThreeOfDelta54}
\rho(\mathrm{A}) = \left(\begin{array}{ccc}0&1&0\\0&0&1\\1&0&0\end{array}\right) \;,\quad \rho(\mathrm{B}) = \left(\begin{array}{ccc}1&0&0\\0&\omega&0\\0&0&\omega^2\end{array}\right) \;,\quad \rho(\mathrm{C}) = \left(\begin{array}{ccc}-1&0&0\\0&0&-1\\0&-1&0\end{array}\right)\;,
\end{equation}
where $\omega:=\exp\nicefrac{2\pi\I}{3}$. 
Since the group of outer automorphisms\footnote{Using GAP~\cite{GAP4} we obtain 
$\mathrm{Aut}\left(\Delta(54)\right) \cong [432, 734]$ and 
$\mathrm{Inn}\left(\Delta(54)\right) \cong [18, 4]$ being the generalized dihedral group.} 
\begin{equation}
\mathrm{Out}\left(\Delta(54)\right) ~\cong~ \nicefrac{\mathrm{Aut}\left(\Delta(54)\right)}{\mathrm{Inn}\left(\Delta(54)\right)} ~\cong~ S_4
\end{equation}
is nontrivial, there are outer automorphisms that might in principle lead to eclectic 
extensions of the traditional flavor symmetry $\Delta(54)$. It turns out that $\Delta(54)$ can be 
extended only in two ways: either by the finite modular group $\Gamma_3'\cong T'$ in the case 
without $\mathcal{CP}$ or by $\Gamma_3'^*\cong\mathrm{SL}(2,3)$ in the case with $\mathcal{CP}$. 
The details are given in the following:

\paragraph{\boldmath Finite modular symmetry $\Gamma'_3$.\unboldmath} We can choose the 
outer automorphisms of $\Delta(54)$
\begin{subequations}
\label{eqs:Delta54outomorphisms1}
\begin{align}
u_\mathrm{S}(\mathrm{A}) & ~=~ \mathrm{B}^2\;, & u_\mathrm{T}(\mathrm{A}) & ~=~ \mathrm{B}\,\mathrm{A}\,\mathrm{B}\;,\phantom{qqqqqqqqqqqqqqqqqqqqqqqqqqqqqqqqqqq}\\
u_\mathrm{S}(\mathrm{B}) & ~=~ \mathrm{A}\;,   & u_\mathrm{T}(\mathrm{B}) & ~=~ \mathrm{B}\;,\\
u_\mathrm{S}(\mathrm{C}) & ~=~ \mathrm{C}\;,   & u_\mathrm{T}(\mathrm{C}) & ~=~ \mathrm{C}\;.
\end{align}
\end{subequations}
Using the presentation eq.~\eqref{eq:GammaN}, one can verify that $u_\mathrm{S}$ and $u_\mathrm{T}$ 
generate the finite modular group $\Gamma'_3 \cong T' \cong \mathrm{SL}(2,3)$. For the 
three-dimensional representation eq.~\eqref{eq:ThreeOfDelta54} of $\Delta(54)$ these outer 
automorphisms can be written by conjugation with\footnote{Note the change of convention compared to ref.~\cite{Baur:2019iai}.}
\begin{equation}\label{eq:ThreeOfTPrime}
\rho(\mathrm{S}) ~=~ -\frac{\I}{\sqrt{3}}\left(\begin{array}{ccc}1&1&1\\1&\omega&\omega^2\\1&\omega^2&\omega\end{array}\right) \quad\mathrm{and}\quad
\rho(\mathrm{T}) ~=~ \left(\begin{array}{ccc}\omega&0&0\\0&1&0\\0&0&1\end{array}\right)\;,
\end{equation}
see eq.~\eqref{eq:ModularFlavorConstraint2}, where possible phases have been fixed by demanding 
that $\rho(\mathrm{S})$ and $\rho(\mathrm{T})$ generate the same finite modular group 
$\Gamma'_3 \cong T' \cong \mathrm{SL}(2,3)$ as $u_\mathrm{S}$ and $u_\mathrm{T}$. By inspecting the 
character table of $T'$, it is easy to see that eq.~\eqref{eq:ThreeOfTPrime} corresponds to the 
$\rep{1}\,\oplus\,\rep{2}''$ irreducible representations of $T'$. Furthermore, we note that 
$(\rho(\mathrm{S}))^2 = \rho(\mathrm{C})$ and the three-dimensional reducible representation 
eq.~\eqref{eq:ThreeOfTPrime} of $T'$ is an irreducible representation of the eclectic flavor 
group generated by $\rho(\mathrm{A})$, $\rho(\mathrm{B})$, $\rho(\mathrm{S})$ and 
$\rho(\mathrm{T})$, being
\begin{equation}
\Omega(1) ~\cong~ [648, 533]\;,
\end{equation}
see ref.~\cite{Jurciukonis:2017mjp} for the definition of $\Omega(1)$ and refs.~\cite{Yao:2015dwa,King:2016pgv} 
for flavor model building based on this symmetry group. Following the 
discussion in section~\ref{sec:LocalFlavorUnification}, the eclectic flavor group $\Omega(1)$ 
is broken spontaneously to various unified flavor subgroups at different points in $\tau$ moduli 
space: 
\begin{subequations}
\begin{align}
\mathrm{generic\ point}                                     &:\;\;\Omega(1)~\rightarrow~ \Delta(54)\;         \hskip1.52cm\cong~ [54, 8] \\
\langle \tau\rangle = \I                                    &:\;\;\Omega(1)~\rightarrow~ \Sigma(36\times 3)\, \hskip9.9mm \cong~ [108, 15]\\
\langle \tau\rangle = \exp\left(\nicefrac{2\pi\I}{3}\right) &:\;\;\Omega(1)~\rightarrow~ \tilde{Y}(0)\;       \hskip1.71cm\cong~ [162, 10]
\end{align}
\end{subequations}
where the flavor groups are defined as follows: 
\begin{enumerate}
\item[i)] $\Delta(54)$ is generated by $\rho(\mathrm{A})$, $\rho(\mathrm{B})$ and $\rho(\mathrm{C})$, 
\item[ii)] $\Sigma(36\times 3)\cong[108, 15]$ is generated by $\rho(\mathrm{A})$, 
$\rho(\mathrm{B})$ and $\rho(\mathrm{S})$ (see ref.~\cite{Hagedorn:2013nra}), and, finally, 
\item[iii)] $\tilde{Y}(0)\cong[162, 10]$ is generated by $\rho(\mathrm{A})$, 
$\rho(\mathrm{B})$, $\rho(\mathrm{C})$ and $\rho(\mathrm{S}\,\mathrm{T})$ (see 
ref.~\cite{Jurciukonis:2017mjp}). 
\end{enumerate}
As a remark, all of these groups have three-dimensional irreducible representations.

\paragraph{\boldmath $\mathcal{CP}$-like modular extension.\unboldmath} 
Following the discussion of section~\ref{sec:CP}, it is possible to extend the identified 
modular symmetry $T'$ with a $\mathcal{CP}$-like transformation $\mathrm K_*$ based on the 
$\Delta(54)$ outer automorphism
\begin{equation}
\label{eq:Delta54outomorphisms2}
u_{\mathrm{K}_*}(\mathrm{A})  ~=~ \mathrm{A}\;,  \qquad
u_{\mathrm{K}_*}(\mathrm{B})  ~=~ \mathrm{B}^2\;,\qquad
u_{\mathrm{K}_*}(\mathrm{C})  ~=~ \mathrm{C}
\end{equation}
that satisfies the conditions~\eqref{eq:ModularFromAutomorphismsWithK*} with $u_{\mathrm S}$ and $u_{\mathrm T}$
given by eqs.~\eqref{eqs:Delta54outomorphisms1}.
In this case, one can see from eq.~\eqref{eq:Definition-uK*} that the representation $\rho(\mathrm K_*)$
of eq.~\eqref{eq:K*trafo} relating the field $\psi(x)$ in the $\rep3$ representation of $\Delta(54)$ with its
conjugate $\overline\psi(x_P)$ in the $\crep3$ representation is given by
\begin{equation}
  \label{eq:K*inDelta54}
  \rho(\mathrm K_*) ~=~ \Id_{3\x3}\,.
\end{equation}
Since $\mathrm K_*$ relates $\psi(x)$ and $\overline\psi(x_P)$, it is more convenient to rewrite
$\mathrm S$, $\mathrm T$ and $\mathrm K_*$ in the $\rep3\oplus\crep3$ representation as
\begin{equation}\label{eq:SG23withDelta54}
\rho_{\rep{6}}(\mathrm{S})   = \left(\!\!\!\begin{array}{cc}\rho(\mathrm S)&0\\0&\rho(\mathrm S)^*\end{array}\!\!\!\right)\!,~
\rho_{\rep{6}}(\mathrm{T})   = \left(\!\!\!\begin{array}{cc}\rho(\mathrm T)&0\\0&\rho(\mathrm T)^*\end{array}\!\!\!\right)\!,~
\rho_{\rep{6}}(\mathrm{K}_*) = \left(\!\!\!\begin{array}{cc}0&\rho(\mathrm K_*)\\ \rho(\mathrm K_*)^*&0\end{array}\!\!\!\right)\!,
\end{equation}
where $\rho(\mathrm S)$ and $\rho(\mathrm T)$ are given by eq.~\eqref{eq:ThreeOfTPrime} and
$\rho(\mathrm K_*)$ by eq.~\eqref{eq:K*inDelta54}. One can easily verify that 
$\rho_{\rep{6}}(\mathrm{S}), \rho_{\rep{6}}(\mathrm{T})$ and $\rho_{\rep{6}}(\mathrm{K}_*)$ 
generate the finite modular group $\Gamma'^*_3\cong\,\mathrm{GL}(2,3)$ and the resulting 
eclectic flavor group is [1296, 2891].

In extra-dimensional models, the traditional flavor symmetry $\Delta(54)$ can originate from 
strings on a $\mathbbm{T}^2/\Z{3}$ orbifold~\cite{Kobayashi:2006wq,Beye:2014nxa}. As shown in 
refs.~\cite{Baur:2019kwi,Baur:2019iai}, $\Delta(54)$ is accompanied nontrivially in this setting 
by a $T'$ or a $\mathrm{GL}(2,3)$ modular symmetry, depending on whether the $\mathcal{CP}$-like 
transformation $\mathrm{K}_*$ is taken into account, in full agreement with our discussion here, see 
also ref.~\cite{Nilles:2018wex}.

\section{Results}
\label{sec:results}

\begin{table}[b!]
\center
\begin{tabular}{|cc|c|l|c|c|}
\hline
flavor group              & GAP        & $\mathrm{Aut}(\mathcal{G}_\mathrm{fl})$ & \multicolumn{2}{c|}{finite modular}          & eclectic flavor\\
$\mathcal{G}_\mathrm{fl}$ & ID         &                                         & \multicolumn{2}{c|}{groups}                  & group\\
\hline
\hline
$Q_8$                     & [ 8, 4 ]   & $S_4$                                   & without $\mathcal{CP}$ & $S_3$               & $\mathrm{GL}(2,3)$\\
\cline{4-6}
                          &            &                                         & with $\mathcal{CP}$    & --                  & --\\
\hline
$\Z3\x\Z3$                & [ 9, 2 ]   & $\mathrm{GL}(2,3)$                      & without $\mathcal{CP}$ & $S_3$               & $\Delta(54)$\\
\cline{4-6}
                          &            &                                         & with $\mathcal{CP}$    & $S_3\times\Z{2}$    & [108, 17]\\
\hline
$A_4$                     & [ 12, 3 ]  & $S_4$                                   & without $\mathcal{CP}$ & $S_3$               & $S_4$\\
                          &            &                                         &                        & $S_4$               & $S_4$\\
\cline{4-6}
                          &            &                                         & with $\mathcal{CP}$    & --                  & --\\
\hline
$T'$                      & [ 24, 3 ]  & $S_4$                                   & without $\mathcal{CP}$ & $S_3$               & $\mathrm{GL}(2,3)$\\
\cline{4-6}
                          &            &                                         & with $\mathcal{CP}$    & --                  & --\\
\hline
$\Delta(27)$              & [ 27, 3 ]  & [ 432, 734 ]                            & without $\mathcal{CP}$ & $S_3$               & $\Delta(54)$\\
                          &            &                                         &                        & $T'$                & $\Omega(1)$\\
\cline{4-6}
                          &            &                                         & with $\mathcal{CP}$    & $S_3\times\Z{2}$    & [108, 17]\\
                          &            &                                         &                        & $\mathrm{GL}(2,3)$  & [1296, 2891]\\
\hline
$\Delta(54)$              & [ 54, 8  ] & [ 432, 734 ]                            & without $\mathcal{CP}$ & $T'$                & $\Omega(1)$\\
\cline{4-6}
                          &            &                                         & with $\mathcal{CP}$    & $\mathrm{GL}(2,3)$  & [1296, 2891]\\
\hline
\end{tabular}
\caption{Examples of traditional flavor groups, their extensions by finite modular groups and the 
resulting eclectic flavor groups. For details, see appendix~\ref{sec:furtherexamples}. 
$\mathrm{Aut}(\mathcal{G}_\mathrm{fl})$ denotes the group of automorphisms of the traditional
flavor group $\mathcal{G}_\mathrm{fl}$.
\label{tab:Examples}}
\end{table}

We have selected a representative set of traditional flavor symmetries that have been used in model 
building. These include
\begin{quote}
$S_3$,\, $Q_8$,\, $\Z3\x\Z3$,\, $A_4$,\, $S_3\times\Z{2}$,\, $T_7$,\, $S_4$,\, $T'$,\, $\Delta(27)$,\,
$\Z9\rtimes\Z3\cong[27, 4]$,
$\mathrm{SL}(2,4)$,\, $\Delta(54)$,\, $A_5$,\, $\Sigma(36\times 3)\cong[108,15]$,\,
$\Sigma(168)\cong[168,42]$ and $\Sigma(72\times 3)\cong[216,88]$.
\end{quote}
This list is not exhaustive, but covers the most promising traditional flavor symmetries considered 
so far. More examples could be added and analyzed upon request. We then classify for each group all 
nontrivial extensions by finite modular symmetries that fulfill the restrictions discussed in 
section~\ref{sec:combining}. This then allows the identification of allowed eclectic flavor groups 
that could be obtained in a bottom-up procedure. Our results are listed in table~\ref{tab:Examples}. 
Surprisingly it turns out that many prominent traditional flavor symmetries, such as $S_3$, 
$D_{12}\cong S_3\x\Z2$, $T_7$, $S_4$, $\Z9\rtimes\Z3$, $\mathrm{SL}(2,4)$ and $A_5$, do not allow 
for any nontrivial modular extension and thus are not included in table~\ref{tab:Examples}. Among 
the groups we have studied, only
\begin{quote}
$Q_8$,\, $\Z3\x\Z3$,\, $A_4$,\, $T'$,\, $\Delta(27)$\, and\, $\Delta(54)$ 
\end{quote}
allow for eclectic extensions and local flavor unification. Furthermore, among those, only the 
traditional flavor groups
\begin{quote}
$\Z3\x\Z3$,\; $\Delta(27)$\; and\; $\Delta(54)$ 
\end{quote}
admit a $\mathcal{CP}$-like transformation in the eclectic flavor group.

This is a quite restrictive situation. A nontrivial extension of the traditional flavor group (as 
required from the top-down argumentation) is limited to just a few specific cases. Flavor model 
building should thus be based on very few examples with an eclectic flavor group that includes a 
nontrivial finite modular group. Since our list of traditional flavor groups is not exhaustive, it
remains to be seen in future work whether more nontrivial cases can be found. Traditional flavor 
groups with a sizable group of outer automorphism are particularly suited for an eclectic extension.

\section{Conclusions}
\label{sec:conclusions}

In the present paper we made an effort to match the bottom-up (BU) and top-down (TD) approaches of 
flavor models based on finite discrete modular symmetries. Up to now, the BU-approach considered 
finite modular groups $\Gamma_N$, where some of the quarks and leptons transform as 
(irreducible) triplet or 
nontrivial singlet representations of $\Gamma_N$. This led to an excellent description of the 
flavor structure of the lepton sector. Efforts towards an ultra-violet completion (TD approach) 
were based on string theory. Up to now only few explicit TD models have been constructed. 
The analysis within string theory, however, leads to a general qualitative picture with the clear 
message that finite modular symmetries do not appear in isolated form, but are accompanied by a 
traditional (non-modular) flavor group. This then leads us to the concept of ``Eclectic Flavor 
Groups'' as a nontrivial product of traditional flavor symmetry and finite modular symmetry.

Given this observation, we might now reconsider the BU-approach and classify candidates for 
eclectic flavor groups from bottom-up. Surprisingly, the number of these candidates turns out to be 
very small. Only a few examples are known (see table~\ref{tab:Examples}). This is the main result 
of the present paper. The fact that we cannot disentangle traditional flavor symmetries and modular 
symmetries is consistent with the picture of ``Local Flavor Unification'', where we find an enhanced 
symmetry at specific regions in moduli space. This would naturally allow different flavor 
structures for quarks and leptons, where quarks (leptons) are predominantly described by 
traditional (modular) flavor groups.

It should be stressed that the concept of eclectic flavor groups is more predictive than the 
consideration of modular symmetries alone. Terms allowed by the modular group might be forbidden by 
the selection rules of the traditional flavor group. In the $\Delta(54)$-example discussed in 
section~\ref{sec:Delta54Example}, four independent trilinear superpotential couplings (of 
$\rep{2}''\oplus\rep{1}$ representations of $T'$) allowed by the $T'$ modular symmetry are reduced 
to a single one due to the presence of $\Delta(54)$~\cite{Nilles:2020}. The enhanced restrictions 
from eclectic flavor groups are especially relevant for the form of the K\"ahler potential. In 
ref.~\cite{Chen:2019ewa} it was pointed out that general terms in the K\"ahler potential reduce the 
predictivity of models based on finite modular symmetries. This problem could be solved within the 
eclectic flavor picture with more restrictions on the K\"ahler potential due to the nontrivial 
combination of finite modular groups and traditional flavor groups.

\subsection*{Acknowledgments}

We are thankful to Andreas Trautner for useful observations.
H.P.N. is supported by the Excellence Cluster ORIGINS, funded by the Deutsche Forschungsgemeinschaft 
(DFG, German Research Foundation) under Germany's Excellence Strategy -- EXC-2094 -- 390783311. The 
work of S.R.-S.\ was partly supported by DGAPA-PAPIIT grant IN100217, CONACyT grants F-252167 and 
278017, PIIF grant and the TUM August--Wilhelm Scheer Program. The work of P.V. is supported by the 
Deutsche Forschungsgemeinschaft (SFB1258).

\appendix

\section{Remarks}
\label{app:remarks}

In this appendix, we comment on some inaccuracies in the literature on modular symmetries in model 
building. First, it is important to note that the modular $\mathrm{S}$ transformation is, in 
general, not of order 2, even though $\mathrm{S}^2$ acts trivially on $\tau$. In detail, from 
eq.~\eqref{eq:ModularTrafoOfFields} we get
\begin{subequations}
\begin{eqnarray}
\tau & \stackrel{\mathrm{S}}{\longmapsto}~ -\frac{1}{\tau}\;,\quad & \psi ~\stackrel{\mathrm{S}}{\longmapsto}~      (-\tau)^{-k} \,\rho(\mathrm{S})\,\psi     \quad\mathrm{for}\quad \mathrm{S} ~=~ \left(\begin{array}{cc} 0 & 1\\ -1 & 0\end{array}\right)\;,\\
\tau & \stackrel{\mathrm{S}^{-1}}{\longmapsto}~           -\frac{1}{\tau}\;,\quad & \psi ~\stackrel{\mathrm{S}^{-1}}{\longmapsto}~ (+\tau)^{-k} \,\rho(\mathrm{S})^{-1}\,\psi\quad\mathrm{for}\quad \mathrm{S}^{-1} ~=~ \left(\begin{array}{cc} 0 & -1\\ 1 & 0\end{array}\right)\;.
\end{eqnarray}
\end{subequations}
Then, $\mathrm{S}^2$ acts as
\begin{subequations}
\begin{eqnarray}
\tau & \stackrel{\mathrm{S}}{\longmapsto} & -\frac{1}{\tau} ~\stackrel{\mathrm{S}}{\longmapsto}~ \tau\;,\\
\psi & \stackrel{\mathrm{S}}{\longmapsto} & (-\tau)^{-k} \,\rho(\mathrm{S})\,\psi ~\stackrel{\mathrm{S}}{\longmapsto}~ \left(\frac{1}{\tau}\right)^{-k} (-\tau)^{-k} \,\rho(\mathrm{S})^2\,\psi ~=~ (-1)^{-k} \,\rho(\mathrm{S})^2\,\psi\;,\label{eq:Ssquare}
\end{eqnarray}
\end{subequations}
while
\begin{subequations}
\begin{eqnarray}
\tau & \stackrel{\mathrm{S}}{\longmapsto} & -\frac{1}{\tau} ~\stackrel{\mathrm{S}^{-1}}{\longmapsto}~ \tau\;,\\
\psi & \stackrel{\mathrm{S}}{\longmapsto} & (-\tau)^{-k} \,\rho(\mathrm{S})\,\psi ~\stackrel{\mathrm{S}^{-1}}{\longmapsto}~ \left(\frac{1}{\tau}\right)^{-k} (\tau)^{-k} \,\rho(\mathrm{S})\,\rho(\mathrm{S}^{-1})\,\psi ~=~ \psi\;,\label{eq:SSinverse}
\end{eqnarray}
\end{subequations}
as expected. Let us compare eq.~\eqref{eq:Ssquare} and eq.~\eqref{eq:SSinverse} in some detail. In 
eq.~\eqref{eq:Ssquare}, the transformation by $\rho(\mathrm{S})^2$ is trivial for $\Gamma_N$ but 
can be nontrivial for $\Gamma'_N$. Moreover, the factor $(-1)^{-k}$ is nontrivial for a general 
modular weight $k\in\mathbb{Q}$ that is not even. Consequently, we see that in general $\mathrm{S}$ 
and $\mathrm{S}^{-1}$ are different transformations for matter fields, even though they act 
identically on the $\tau$ modulus.

\section{Explicit Examples}
\label{sec:furtherexamples}

\subsection[Traditional flavor symmetry Q8]{\boldmath Traditional flavor symmetry $Q_8$\unboldmath}

Let us consider the traditional flavor symmetry $Q_8$ of order 8 (GAP ID [8,4]). We choose the 
irreducible two-dimensional representation of $Q_8$ given by~\cite{Ishimori:2010au}
\begin{equation}
\rho(\mathrm{A}) = \left(\begin{array}{ccc}\I&0\\0&-\I\end{array}\right) \;,\quad \rho(\mathrm{B}) = \left(\begin{array}{ccc}0&\I\\\I&0\end{array}\right) \;.
\end{equation}
The full automorphism group of $Q_8$ is $S_4$ with 24 elements. Out of these, we can identify two 
outer automorphisms that generate the finite modular group $S_3$, i.e.\ 
\begin{subequations}
\begin{align}
u_\mathrm{S}(\mathrm{A}) & ~=~ \mathrm{A}^3\;,           & u_\mathrm{T}(\mathrm{A}) & ~=~ \mathrm{B}\;,\phantom{qqqqqqqqqqqqqqqqqqqqqqqqqqqqqqqqqqq}\\
u_\mathrm{S}(\mathrm{B}) & ~=~ \mathrm{A}\,\mathrm{B}\;, & u_\mathrm{T}(\mathrm{B}) & ~=~ \mathrm{A}\;.
\end{align}
\end{subequations}
Then, there are two choices of matrices $\rho(\mathrm{S})$ and $\rho(\mathrm{T})$ that realize 
these automorphisms via eq.~\eqref{eq:ModularFlavorConstraint2} and generate $S_3$, being
\begin{equation}\label{eq:Q8Representation}
\rho(\mathrm{S}) = \alpha\,\left(\begin{array}{ccc}0&\exp\nicefrac{2\pi\I}{8}\\\exp\nicefrac{-2\pi\I}{8}&0\end{array}\right) \;,\quad \rho(\mathrm{T}) = \frac{\alpha}{\sqrt{2}}\,\left(\begin{array}{ccc}-1&-1\\-1&1\end{array}\right)\;,
\end{equation}
for $\alpha = \pm 1$. For both choices of $\alpha$, $\rho(\mathrm{S})$ and $\rho(\mathrm{T})$ build 
a doublet of $S_3$. Moreover, the eclectic flavor group, generated by $\rho(\mathrm{A})$, 
$\rho(\mathrm{B})$, $\rho(\mathrm{S})$ and $\rho(\mathrm{T})$, turns out to be $\mathrm{GL}(2,3)$.
It is interesting to note that the $Q_8$ traditional flavor symmetry does not allow for an 
eclectic extension with $\mathcal{CP}$.

\subsection[Traditional flavor symmetry Z3 x Z3]{\boldmath Traditional flavor symmetry $\Z{3}\times\Z{3}$\unboldmath}

We choose a (reducible) three-dimensional representation of $\Z{3}\times\Z{3}$ given by
\begin{equation}\label{eq:Z3xZ3part1}
\rho(\mathrm{A}) = \left(\begin{array}{ccc}\omega&0&0\\0&1&0\\0&0&\omega^2\end{array}\right) \;,\quad 
\rho(\mathrm{B}) = \left(\begin{array}{ccc}1&0&0\\0&\omega&0\\0&0&\omega^2\end{array}\right) \;.
\end{equation}
The full automorphism group of $\Z{3}\times\Z{3}$ is $\mathrm{GL}(2,3)$ with 48 elements. Since the 
group of inner automorphisms of $\Z{3}\times\Z{3}$ is trivial, all elements of $\mathrm{GL}(2,3)$ 
are outer automorphisms. It turns out that there are two classes of outer automorphisms that 
generate finite modular groups, either without or with $\mathcal{CP}$:
\begin{itemize}
\item[i)]Without $\mathcal{CP}$, we can choose the outer automorphisms
\begin{subequations}
\label{eq:exampleS3fromZ3xZ3}
\begin{align}
u_\mathrm{S}(\mathrm{A}) & ~=~ \mathrm{A}^2\;,           & u_\mathrm{T}(\mathrm{A}) & ~=~ \mathrm{B}\;,\phantom{qqqqqqqqqqqqqqqqqqqqqqqqqqqqqqqqqqq}\\
u_\mathrm{S}(\mathrm{B}) & ~=~ \mathrm{A}^2\mathrm{B}\;, & u_\mathrm{T}(\mathrm{B}) & ~=~ \mathrm{A}\;,
\end{align}
\end{subequations}
which generate the finite modular group $S_3$. One possibility to realize these outer 
automorphisms via conjugation with matrices $\rho(\mathrm{S})$ and $\rho(\mathrm{T})$ is given by
the choice
\begin{equation}\label{eq:Z3xZ3part2}
\rho(\mathrm{S}) = \left(\begin{array}{ccc}0&0&-\I\\0&-1&0\\ \I&0&0\end{array}\right) \;,\quad 
\rho(\mathrm{T}) = \left(\begin{array}{ccc}0&-1&0\\-1&0&0\\0&0&-1\end{array}\right)\;,
\end{equation}
see eq.~\eqref{eq:ModularFlavorConstraint2}. This three-dimensional representation of $S_3$ 
decomposes into a $\rep{2}\oplus\rep{1}'$. Interestingly, the eclectic flavor group, generated 
by $\rho(\mathrm{A})$, $\rho(\mathrm{B})$, $\rho(\mathrm{S})$ and $\rho(\mathrm{T})$, is 
$\Delta(54)$ and the representation eq.~\eqref{eq:Z3xZ3part1} and eq.~\eqref{eq:Z3xZ3part2} is 
three-dimensional.

\item[ii)]Moreover, when combining eq.~\eqref{eq:exampleS3fromZ3xZ3} with the outer 
automorphism
\begin{equation}
u_{\mathrm{K}_*}(\mathrm{A}) ~=~ \mathrm{A}^2\;,\qquad 
u_{\mathrm{K}_*}(\mathrm{B}) ~=~ \mathrm{B}^2
\end{equation}
we see that the $\mathcal{CP}$-enhanced finite modular group is $\Gamma_2^*\cong S_3\x\Z2$,
where $\mathrm K_*$ can be represented as
\begin{equation}\label{eq:Z3xZ3part3}
\rho(\mathrm{K}_*) = \left(\begin{array}{ccc}1&0&0\\0&1&0\\0&0&-1\end{array}\right) \;,\quad 
\end{equation}
which satisfies eq.~\eqref{eq:ModularFlavorConstraintWithK*}. The action on the matter fields
$(\psi,\overline\psi)^T$ is realized by rewriting all modular generators in the six-dimensional 
representation, as in eq.~\eqref{eq:SG23withDelta54}. In this six-dimensional representation, one 
can easily confirm using GAP that the eclectic flavor group including $\mathcal{CP}$ is [108,17].
\end{itemize}

\subsection[Traditional flavor symmetry A4]{\boldmath Traditional flavor symmetry $A_4$\unboldmath}

The generators of the traditional flavor symmetry $A_4$ (GAP ID [12,3]) can be given in the triplet 
representation by the matrices
\begin{equation}
\label{eq:A4flavorGenerators}
\rho(\mathrm{A}) = \left(\begin{array}{ccc}1&0&0\\0&-1&0\\0&0&-1\end{array}\right) \;,\quad 
\rho(\mathrm{B}) = \left(\begin{array}{ccc}0&1&0\\0&0&1\\1&0&0\end{array}\right)\,.
\end{equation}
The full automorphism group of $A_4$ is $S_4$, which contains only two finite modular groups 
generated by the outer automorphisms:
\begin{enumerate}
\item[i)] Two outer automorphisms that generate the finite modular group $\Gamma_2\cong S_3$ are
\begin{subequations}
\label{eq:exampleS3fromA4}
\begin{align}
u_\mathrm{S}(\mathrm{A}) & ~=~ \mathrm{B}^2\,\mathrm{A}\,\mathrm{B}\;,  & u_\mathrm{T}(\mathrm{A}) & ~=~ \mathrm{B}\,\mathrm{A}\,\mathrm{B}^2\;,\phantom{qqqqqqqqqqqqqqqqq}\\
u_\mathrm{S}(\mathrm{B}) & ~=~ \mathrm{B}^2\;,                          & u_\mathrm{T}(\mathrm{B}) & ~=~ \mathrm{B}^2\;.
\end{align}
\end{subequations}
In the representation~\eqref{eq:A4flavorGenerators}, these automorphisms can be expressed as
\begin{equation}\label{eq:exampleS3fromA4matrices}
\rho(\mathrm{S}) = \left(\begin{array}{ccc}0&1&0\\1&0&0\\0&0&1\end{array}\right) \;,\quad 
\rho(\mathrm{T}) = \left(\begin{array}{ccc}0&0&1\\0&1&0\\1&0&0\end{array}\right)\;.
\end{equation}
The generators $\rho(\mathrm{A})$, $\rho(\mathrm{B})$, $\rho(\mathrm{S})$ and $\rho(\mathrm{T})$ 
generate the eclectic flavor group $S_4$, which is isomorphic to the full automorphism group $S_4$.

\item[ii)] Two outer automorphisms that generate the finite modular group $\Gamma_4\cong S_4$ are
\begin{subequations}
\label{eq:exampleS4fromA4}
\begin{align}
u_\mathrm{S}(\mathrm{A}) & ~=~ \mathrm{B}^2\,\mathrm{A}\,\mathrm{B}\;,  & u_\mathrm{T}(\mathrm{A}) & ~=~ \mathrm{B}\,\mathrm{A}\,\mathrm{B}^2\;,\phantom{qqqqqqqqqqqqqqq}\\
u_\mathrm{S}(\mathrm{B}) & ~=~ \mathrm{B}^2\;,                          & u_\mathrm{T}(\mathrm{B}) & ~=~ \mathrm{B}\,\mathrm{A}\,\mathrm{B}\;.
\end{align}
\end{subequations}
In the representation~\eqref{eq:A4flavorGenerators}, these automorphisms can be expressed as
\begin{equation}\label{eq:exampleS4fromA4matrices}
\rho(\mathrm{S}) = \left(\begin{array}{ccc}0&1&0\\1&0&0\\0&0&1\end{array}\right) \;,\quad 
\rho(\mathrm{T}) = \left(\begin{array}{ccc}0&0&1\\0&-1&0\\-1&0&0\end{array}\right)\;,
\end{equation}
and the eclectic extension of the traditional flavor symmetry $A_4$ by the $\Gamma_4\cong S_4$ 
finite modular group yields the eclectic flavor group $S_4$.
\end{enumerate}
Let us conclude this example with two remarks, both related to the fact that the group of outer 
automorphisms of $A_4$ is $\Z2$, i.e.\ very small. First, the $A_4$ traditional flavor group does 
not allow for an eclectic flavor group with $\mathcal{CP}$. Secondly, for both eclectic extensions 
of the traditional flavor group $A_4$ by either by $S_3$ or $S_4$ the eclectic flavor group is 
$S_4$. Note that, even though the eclectic flavor groups are identical (i.e.\ $S_4$), the eclectic 
extensions of $A_4$ by $S_3$ or $S_4$ yield different theories, since, for example, Yukawa 
couplings have to be modular forms of either $S_3$ or $S_4$.

\subsection[Traditional flavor symmetry T']{\boldmath Traditional flavor symmetry $T'$\unboldmath}

The generators of the traditional flavor symmetry $T'$ (GAP ID [24,3]) can be given in the 
$\rep1\oplus\rep2$ representation by the matrices
\begin{equation}
\label{eq:T'flavorGenerators}
\rho(\mathrm{A}) = \left(\begin{array}{ccc}1&0&0\\0&\omega^2&0\\0&0&\omega\end{array}\right)\!,~ 
\rho(\mathrm{B}) = \left(\begin{array}{ccc}1&0&0\\0&-1&0\\0&0&-1\end{array}\right)\!,~
\rho(\mathrm{C}) = \frac{-\I}{\sqrt3}\left(\begin{array}{ccc}\I\sqrt3&0&0\\0&1&\sqrt2\\0&\sqrt2&-1\end{array}\right)\!,
\end{equation}
where $\omega:=\exp\nicefrac{2\pi\I}{3}$. The full automorphism group of $T'$ is $S_4$. One can 
choose the following two outer automorphisms of this group to generate the finite 
modular group $\Gamma_2\cong S_3$:
\begin{subequations}
\label{eq:exampleS3fromT'}
\begin{align}
u_\mathrm{S}(\mathrm{A}) & ~=~ \mathrm{A}^2\;,                                     & u_\mathrm{T}(\mathrm{A}) & ~=~ \mathrm{A}^2\;,\phantom{qqqqqqqqqqqqqqqqq}\\
u_\mathrm{S}(\mathrm{B}) & ~=~ \mathrm{B}\;,                                       & u_\mathrm{T}(\mathrm{B}) & ~=~ \mathrm{B}\;,\\
u_\mathrm{S}(\mathrm{C}) & ~=~ \mathrm{C}\,\mathrm{A}\,\mathrm{C}\,\mathrm{A}^2\;, & u_\mathrm{T}(\mathrm{C}) & ~=~ \mathrm{C}\,\mathrm{B}\;.
\end{align}
\end{subequations}
In the representation~\eqref{eq:T'flavorGenerators}, these automorphisms are given by
\begin{equation}\label{eq:exampleS3fromT'matrices}
\rho(\mathrm{S}) = \left(\begin{array}{ccc}\alpha&0&0\\0&0&\beta\,\I\,\omega\\0&-\beta\,\I\,\omega^2&0\end{array}\right) \;,\quad 
\rho(\mathrm{T}) = \left(\begin{array}{ccc}\alpha&0&0\\0&0&\beta\,\I\\0&-\beta\,\I&0\end{array}\right)\;,
\end{equation}
for $\alpha=\pm 1$ and $\beta=\pm 1$. The generators $\rho(\mathrm{A})$, $\rho(\mathrm{B})$, 
$\rho(\mathrm{C})$, $\rho(\mathrm{S})$ and $\rho(\mathrm{T})$ generate the eclectic flavor 
group $\mathrm{GL}(2,3)$. 

Note that the $T'$ traditional flavor symmetry does not allow for an eclectic flavor group with 
$\mathcal{CP}$.

\subsection[Traditional flavor symmetry Delta(27)]{\boldmath Traditional flavor symmetry $\Delta(27)$\unboldmath}

Next, we choose the traditional flavor symmetry $\Delta(27)$ (GAP ID [27, 3]). It can be generated 
in a triplet representation by the matrices
\begin{equation}
\label{eq:Delta27flavorGenerators}
\rho(\mathrm{A}) = \left(\begin{array}{ccc}0&1&0\\0&0&1\\1&0&0\end{array}\right) \;,\quad 
\rho(\mathrm{B}) = \left(\begin{array}{ccc}\omega&0&0\\0&1&0\\0&0&\omega^2\end{array}\right)\;,\quad 
\rho(\mathrm{C}) = \left(\begin{array}{ccc}\omega^2&0&0\\0&\omega&0\\0&0&1\end{array}\right)\;,
\end{equation}
where $\rho(\mathrm{C}) = \rho(\mathrm{A})^2\,\rho(\mathrm{B})\,\rho(\mathrm{A})$, see 
ref.~\cite{Ishimori:2010au}. Out of the full group of automorphisms with GAP ID [432, 734] we 
identify two classes of outer automorphisms that generate finite modular groups. 
\begin{enumerate}
\item[i)] We can choose
\begin{subequations}
\label{eq:exampleS3fromDelta27}
\begin{align}
u_\mathrm{S}(\mathrm{A}) & ~=~ \mathrm{A}^2\;, & u_\mathrm{T}(\mathrm{A}) & ~=~ \mathrm{A}^2\;,\phantom{qqqqqqqqqqqqqqqqqqqqqqqqqqqqqqqqqqq}\\
u_\mathrm{S}(\mathrm{B}) & ~=~ \mathrm{B}^2\;, & u_\mathrm{T}(\mathrm{B}) & ~=~ \mathrm{A}^2\,\mathrm{B}^2\,\mathrm{A}\;.
\end{align}
\end{subequations}
These automorphisms generate $S_3$. Then, there are two choices of matrices that 
generate $S_3$ and realize these outer automorphisms by conjugation, 
eq.~\eqref{eq:ModularFlavorConstraint2}, being
\begin{equation}
\rho(\mathrm{S}) = \alpha\,\left(\begin{array}{ccc}0&0&1\\0&1&0\\1&0&0\end{array}\right) \;,\quad 
\rho(\mathrm{T}) = \alpha\,\left(\begin{array}{ccc}1&0&0\\0&0&1\\0&1&0\end{array}\right)\;,
\end{equation}
for $\alpha = \pm 1$. In both cases, $\rho(\mathrm{S})$ and $\rho(\mathrm{T})$ generate the finite 
modular group $\Gamma_2 \cong S_3$, where the triplet of $\Delta(27)$ builds a 
$\rep{2} \oplus\rep{1}$ of $S_3$ if $\alpha=1$, or $\rep{2} \oplus\rep{1}'$ if $\alpha=-1$. In both 
cases, the eclectic flavor group is $\Delta(54)$.

Incorporating the $\mathcal{CP}$-like transformation $\mathrm K_*$ can be done by including the 
outer automorphism
\begin{equation}
u_{\mathrm{K}_*}(\mathrm{A}) ~=~ \mathrm{A}\;,\qquad 
u_{\mathrm{K}_*}(\mathrm{B}) ~=~ \mathrm{B}\;\mathrm A^2\;\mathrm B\,,
\end{equation}
which, together with eq.~\eqref{eq:exampleS3fromDelta27}, leads to the $\mathcal{CP}$-enhanced 
finite modular group $S_3\x\Z2$. In the representation eq.~\eqref{eq:Delta27flavorGenerators}, it 
can be written as
\begin{equation}
\rho(\mathrm{K}_*) = \frac{\I}{\sqrt3}\left(\begin{array}{ccc}1&\omega^2&\omega^2\\\omega^2&1&\omega^2\\\omega^2&\omega^2&1\end{array}\right) \;,\quad 
\end{equation}
which satisfies eq.~\eqref{eq:ModularFlavorConstraintWithK*}. The action on the matter fields 
$(\psi,\overline\psi)^T$ is realized by rewriting all modular generators in the six-dimensional 
representation, as in eq.~\eqref{eq:SG23withDelta54}. In this case, the eclectic flavor group 
including $\mathcal{CP}$ is [108,17].

\item[ii)] Furthermore, we can choose
\begin{subequations}
\label{eq:exampleGL23fromDelta27}
\begin{align}
u_\mathrm{S}(\mathrm{A}) & ~=~ \mathrm{B}^2\,\mathrm{A}\;,   & u_\mathrm{T}(\mathrm{A}) & ~=~ \mathrm{B}\,\mathrm{A}\;,\phantom{qqqqqqqqqqqqqqqqqqqqqqqqqqqqqq}\\
u_\mathrm{S}(\mathrm{B}) & ~=~ (\mathrm{A}\,\mathrm{B})^2\;, & u_\mathrm{T}(\mathrm{B}) & ~=~ \mathrm{B}\;.
\end{align}
\end{subequations}
These outer automorphisms generate $T'$. Then, there are three choices of matrices 
that generate $T'$ and realize these automorphisms by conjugation, being
\begin{equation}
\rho(\mathrm{S}) = -\frac{\I}{\sqrt{3}}\,\left(\begin{array}{ccc}1&\omega&\omega\\\omega^2&\omega&\omega^2\\\omega^2&\omega^2&\omega\end{array}\right) \;,\quad 
\rho(\mathrm{T}) = \omega^k\,\left(\begin{array}{ccc}1&0&0\\0&\omega^2&0\\0&0&\omega^2\end{array}\right)\;,
\end{equation}
for $k=0,1,2$. In these cases, the eclectic flavor group results in $\Omega(1)$ (with GAP ID [648, 533]).

Incorporating the $\mathcal{CP}$-like transformation $\mathrm K_*$ can be done by including the 
outer automorphism
\begin{equation}
u_{\mathrm{K}_*}(\mathrm{A}) ~=~ \mathrm{A}^2\,\mathrm B\;,\qquad 
u_{\mathrm{K}_*}(\mathrm{B}) ~=~ \mathrm A^2\,\mathrm{B}\,\mathrm A\,,
\end{equation}
which, together with eq.~\eqref{eq:exampleGL23fromDelta27}, leads to the $\mathcal{CP}$-enhanced 
finite modular group $\mathrm{GL}(2,3)$. In the representation eq.~\eqref{eq:Delta27flavorGenerators}, 
it can be written as
\begin{equation}
\rho(\mathrm{K}_*) = e^{\I \gamma}\left(\begin{array}{ccc}\omega^2&0&0\\0&0&1\\0&1&0\end{array}\right)\;,
\end{equation}
which satisfies eq.~\eqref{eq:ModularFlavorConstraintWithK*} with $\gamma\in\mathbbm{R}$. The 
action on the matter fields $(\psi,\overline\psi)^T$ is realized by rewriting all modular 
generators in the six-dimensional representation, as in eq.~\eqref{eq:SG23withDelta54}. The 
resulting eclectic flavor group is [1296, 2891].
\end{enumerate}

\subsection[Trivial extension of Delta(54)]{Trivial extension of the traditional flavor symmetry $\boldsymbol{\Delta(54)}$}

As discussed before section~\ref{sec:eclectic}, inner automorphisms can only produce a trivial 
extension of a traditional flavor symmetry. To illustrate this scenario, let us entertain the 
possibility of $\Delta(54)$ being extended by a $\Gamma_2\cong S_3$ finite modular symmetry 
generated by the $\Delta(54)$ inner automorphisms
\begin{subequations}
\begin{align}
u_\mathrm{S}(\mathrm{A}) & ~=~ \mathrm{A}^2\;, & u_\mathrm{T}(\mathrm{A}) & ~=~ \mathrm{A}^2\;,\phantom{qqqqqqqqqqqqqqqqqqqqqqqqqqqqqqqqqqq}\\
u_\mathrm{S}(\mathrm{B}) & ~=~ \mathrm{B}^2\;, & u_\mathrm{T}(\mathrm{B}) & ~=~ \mathrm{A}^2\mathrm{B}^2\mathrm{A}\;,\\
u_\mathrm{S}(\mathrm{C}) & ~=~ \mathrm{C}\;,   & u_\mathrm{T}(\mathrm{C}) & ~=~ \mathrm{A}\mathrm{C}\;;
\end{align}
\end{subequations}
see section~\ref{sec:Delta54Example}. For the three-dimensional representation 
eq.~\eqref{eq:ThreeOfDelta54} of $\Delta(54)$, these are given in terms of 
eq.~\eqref{eq:ModularFlavorConstraint2} by the matrix representations
\begin{equation}\label{eq:S3withDelta54-ex1}
\rho(\mathrm{S}) ~=~ -\left(\begin{array}{ccc}1&0&0\\0&0&1\\0&1&0\end{array}\right) \quad\mathrm{and}\quad
\rho(\mathrm{T}) ~=~ -\left(\begin{array}{ccc}0&1&0\\1&0&0\\0&0&1\end{array}\right)\;.
\end{equation}
From eq.~\eqref{eq:GammaN}, one can easily see that $\rho(\mathrm{S})$ and $\rho(\mathrm{T})$ 
generate the finite modular group $\Gamma_2 \cong S_3$. Notice however that 
$\rho(\mathrm{S})=\rho(\mathrm{C})$ and $\rho(\mathrm{T})=\rho(\mathrm{A})^2\rho(\mathrm{C})$. As 
in eq.~\eqref{eq:trivialextension}, we observe here that the action of the $S_3$ modular generators 
is compensated by the $\Delta(54)$ elements represented by $\rho(\mathrm{C})^{-1}$ and 
$\rho(\mathrm{A}^2\mathrm{C})^{-1}$. Therefore, the action of the finite modular group $S_3$ based 
on inner automorphisms can be given by the trivial representation $\rho(\gamma)=\Id$, which amounts 
to a trivial extension of the traditional flavor symmetry.

As a side remark, a careful inspection reveals that there are no outer automorphisms of $\Delta(54)$ 
that generate a $\Gamma_2$ finite modular group and simultaneously satisfy eqs.~\eqref{eq:GammaN} and~\eqref{eq:ModularFlavorConstraint2}.
Thus, the $\Delta(54)$ traditional flavor group cannot be enhanced nontrivially by a $\Gamma_2$ finite 
modular group.

\bibliography{Orbifold}
\bibliographystyle{NewArXiv}
\end{document}